\documentclass[manuscript]{aastex}

\usepackage{graphicx}

\usepackage{amsmath}

\usepackage{booktabs}

\title{Dependence of Coronal Loop Heating on the Characteristics of Slow Photospheric Motions}

\shorttitle{Photospheric Motions and Resultant Heating}
\shortauthors{Ritchie et al}

\author{M.L. Ritchie, A.L. Wilmot-Smith, G. Hornig}
\affil{University of Dundee}
\affil{Division of Mathematics, Dundee, DD1 4HN}
\email{mritchie@maths.dundee.ac.uk}
\author{A.L. Wilmot-Smith}
\affil{University of St Andrews}
\affil{Mathematical Institute, St Andrews, KY16 9SS}
\author{G. Hornig}
\affil{University of Dundee}
\affil{Division of Mathematics, Dundee, DD1 4HN}

\begin{document}

%\maketitle
\begin{abstract}The Parker hypothesis (\citet{Parker1972}) assumes that heating of coronal loops occurs due to reconnection, induced when photospheric motions braid field lines to the point of current sheet formation. In this  contribution we address the question of how the nature of photospheric motions affects heating of braided coronal loops. We design a series of boundary drivers and quantify their properties in terms of complexity and helicity injection. We examine a series of long-duration full resistive MHD simulations in which a simulated coronal loop, consisting of initially uniform field lines, is subject to these photospheric flows. Braiding of the loop is continually driven until differences in behaviour induced by the drivers can be characterised. It is shown that heating is crucially dependent on the nature of the photospheric driver - coherent motions typically lead to fewer large energy release events, while more complex motions result in more frequent but less energetic heating events.
\end{abstract}

\section{Introduction}
Finding an explanation for the unexpectedly high temperatures observed in the corona of our Sun has been an area of extensive research for decades in the solar physics community.  Most of the models for coronal heating involve magnetic reconnection, a change in the topology of the magnetic field which allows it to move to a lower energy state, thereby releasing energy into the surrounding plasma. \\

Given the high conductivity of the solar corona, magnetic reconnection can only take place if short length scales (i.e thin current sheets) develop in the field. One theory for the way in which these small scales could develop in coronal loops is via Parker's notion of topological dissipation (\citet{Parker1972}). The idea is that complex but slow photospheric motions acting on simple magnetic fields (with no null points or other topological features) will twist and tangle the strands of the loop. These will then attempt to relax to a force-free state in the low plasma beta environment of the corona. Parker hypothesised that the space of force-free fields is restricted and hence an arbitrarily tangled loop will typically have no smooth force-free equilibrium which can be reached via an ideal relaxation. Instead such a relaxation will develop tangential discontinuities corresponding to singular current sheets. Hence in a real corona with finite resistivity diffusion effects would become large enough for reconnection to occur.\\

A great deal of research has tested the Parker hypothesis, with several works lending support to the idea (e.g \citet{Long1998}; \citet{Janse2009}; \citet{Low2010}). On the other hand \citet{Bineau1972} proved that a smooth force-free field exists for low enough $\alpha$ (where for a force-free field $\nabla \times {\bf B} = \alpha {\bf B}$). Subsequent numerical simulations into the problem (e.g \cite{Balle1988}; \citet{Mikic1989}; \citet{craigsneyd2005}; \citet{Pontin2014}) have suggested that rather than singular current layers forming,  smooth force-free equilibria exist for arbitrary footpoint displacements but that the thickness of current layers decreases exponentially with complexity. Hence, if we assume a photospheric driver, which constantly increases the braiding of coronal fields, current layers will become thin enough to become dynamically important. In either of these cases (the formation of singular currents or those with a thickness exponentially decreasing in time) reconnection will eventually be enabled in a resistive plasma.

The primary question addressed in this paper is how loops respond to driving on very long timescales. Some simple considerations show that this response will depend not only on the coronal loop plasma environment but also on the nature of the photospheric motions. For example, in a corona with a very high resistivity excess magnetic energy will be rapidly dissipated leading to a uniform heating rate. However, in a corona with very low resistivity (or a perfectly conducting corona) then photospheric motions can both inject and remove Poynting flux (\citet{Yeates2014}). In an idealised case with a single vortex rotating in a sequence of opposite directions (left-right-left etc.) there would be no overall energy input to the corona. A more realistic driver will braid the overlying field but can also \lq unbraid' the field, depending on how the motions relate to the overlying field configuration. With a low resistivity dissipation will occur in thin current sheets. Simulations of resistive MHD decay ( e.g \citet{detal2011}) show a cascade effect of current sheets and a decay timescale that depends strongly on resisitivity. The long term evolution will then depend on this timescale together with that of the photospheric motions. 

Several previous simulations have examined the question of long term coronal field evolution. A common feature to all simulations so far (see overview by \citet{ws2015}) is that the loops evolve into statistically steady states where quantities fluctuate intermittently in time about average levels and the Poynting flux and dissipation are decoupled on short timescales (e.g \citet{LongSud1994}; \citet{HendrixHov1996}; \citet{GalNord1996}; \citet{Rapp2007}, 2008, 2010, 2013; \citet{Ng2012}). Simulations by e.g.~\citet{Ng2012} and \citet{Rapp2007}) have shown that the fluctuations have an increasingly intermittent, bursty character at higher magnetic Reynolds number $R_m$. However, a detailed understanding of how the states  depend on the driver properties is lacking. \cite{GalNord1996} applied a shearing profile (changing the shear direction at randomly chosen times) while most other works have applied incompressible models of convection. These flows are cellular rotational profiles that can either be time dependent or independent. Two comparison cases of a single stationary localised vortex and of a stationary shear were taken by \citet{Rapp2010}. Each of these cases had the same statistically steady state behaviour, which would suggest the profile of photospheric driving is not critical for coronal heating. In particular the uniformly twisting driver profile in \cite{Rapp2010} did not, after the initial kink-instability had occurred, lead to significant build-up of twist in the loop or to repeated kink-instabilities as might have been expected. In a contrasting find \citet{wsetal2011} examined the resistive evolution of two braided fields. These fields were imposed as initial conditions rather than being constructed by boundary motions and both were found to undergo a dynamic evolution. In the first case was a complex braid where the magnetic field had no overall helicity, being constructed from three twists of positive and negative sign (modelled on the pigtail braid). The second case was a coherent braid in which the magnetic field had a positive helicity, constructed from six positively twisted regions. The decay of these two braids gave very distinct behaviours, with the complex case leading to a rather homogeneous heating and the coherent case to a very localised heating. 

The aim of this paper is to address in more detail the question of how the nature of photospheric motions changes the type of loop heating. For this we design a time-dependent photospheric driver with adjustable properties and simulate the long time evolution of loops subjected to various realisations of the driver. The available properties of the driver include the level of helicity injection as well as flow complexity, measured by the topological entropy. These properties are discussed and motivated further in Section $2$ with the details of the driver being given in Section $3$. In Section $4$ the results of the simulations are presented and a discussion is given in Section $5$. 

\section{Concept}
The idea of the simulation is to design drivers with varying characteristics which represent photospheric motions, and drive continuously for a long time period in a $3$D MHD resistive environment to determine how the different properties impact on the nature of the heating of the loop. These can be seen as toy drivers - we are concerned with how the fundamental properties of photospheric motions could affect the nature of heating, not with designing the most physically realistic model of photospheric flows. We take a magnetic field in a high aspect ratio box, representing a volume containing a straightened coronal loop. This field is initially uniform and a driver is applied at the lower boundary, moving the footpoints at this end of the simulated photosphere. The typically high plasma beta in the photosphere and largely ideal environment in the corona means that field lines become braided in time as the boundary is continuously driven. We allow our simulations to run long enough to examine how the loop is heated as a result of the different types of flows. Again, at this point we are only interested in comparing the behaviour arising from motions with different properties. What sort of impact does complexity of the drivers have on heating? How much of a factor is the level of helicity in the system? Is it possible to reach a statistically steady state? Does the driving spark instabilities, and how may they be resolved? These are the questions we can address with our simulations. 

\section{Simulation Setup}

\subsection{Boundary Driver}
Our driver consists of two incompressible, so-called blinking vortices lying in the same plane, illustrated schematically in Figure \ref{basic_concept}. The detailed expression for the profile is given in Appendix 1.  One vortex ramps up to the maximum velocity, maintained for $10\pi$ time units before ramping down, at which point a second vortex follows the same process. The twisting motion of one vortex is completed in $12 \pi$ time units, giving a driver period of $24 \pi$. By \lq completed' we mean the vortex has ramped up to its maximum, driven and then ramped down to zero velocity again. The driver is periodic in time and the profile is chosen such that the maximum velocity never exceeds $1/10$ of the Alfv\'en velocity so that the driving can be considered as slow.   \\
In the first instance we consider two categories of driver with this fundamental action. In the first category the two vortices have opposite circulation, while in the second the vortices spin in the same way. These two different types of action have different effects on the braiding of the field. We refer to the opposite case as the the low helicity case: consider an idealised situation with both vortices centred at the origin. One vortex spins and then the other starts to spin the other way. Some of the braiding induced by the first vortex will be undone by the second. Negative helicity is injected into the system, cancelling with the positive helicity from the first spin, and Poynting flux flow is reversed. On the other hand, the equal circulation case is referred to as the high helicity case: the second vortex will only add to the level of Poynting flux into the domain and the positive helicity in the system. \\
Within these two categories we have a further subcategory. We fix one vortex at the origin, and place the second at $(x_0,0)$. The degree of braiding induced in an ideal coronal volume will vary with $x_0$ - for $x_0$ very close to zero, the action of one vortex will affect action carried out by the other more than for $x_0$ much larger than zero, (recall the exponential form), resulting in braids of different complexity. \\
We measure this complexity using the quantity of topological entropy of the flows described by the driver functions. Further details can be found in Appendix $2$, but one description of the quantity given in \citet{Newhousedefn} is the \lq asymptotic growth rate of material lines\rq \ in a flow. Consider the flow generated by our drivers. If we follow the trajectories of particles at some initial positions in this flow, and plot them in time, where time is our z-axis, we obtain a braid. In an idealised situation the field lines anchored in the flow at the same initial positions would be twisted and tangled according to these trajectories. The braid of field lines would mirror the braid of trajectories. The method of calculation of topological entropy involves applying the braiding action to a material line in a flow and measuring the rate of stretching of the line. Therefore by analogy, in our coronal loop scenario, higher topological entropy means a more complex driving motion and more complex tangling of the strands. We find that in general the low helicity case has higher entropy values than the high helicity case. Therefore we can also refer to the case of opposite twists case as the complex case and equal twists as the coherent case. \\
The underlying properties of the $6$ runs for which results are presented here are summarised in Table $1$. Our two main cases are: Group $1$, the opposite twist, complex but low helicity case, within which we vary $x_0$, and Group $2$, the equal twist, coherent and high helicity situation where we test the same values of $x_0$. \\

\begin{figure}[!h]
\includegraphics[width=0.5\textwidth]{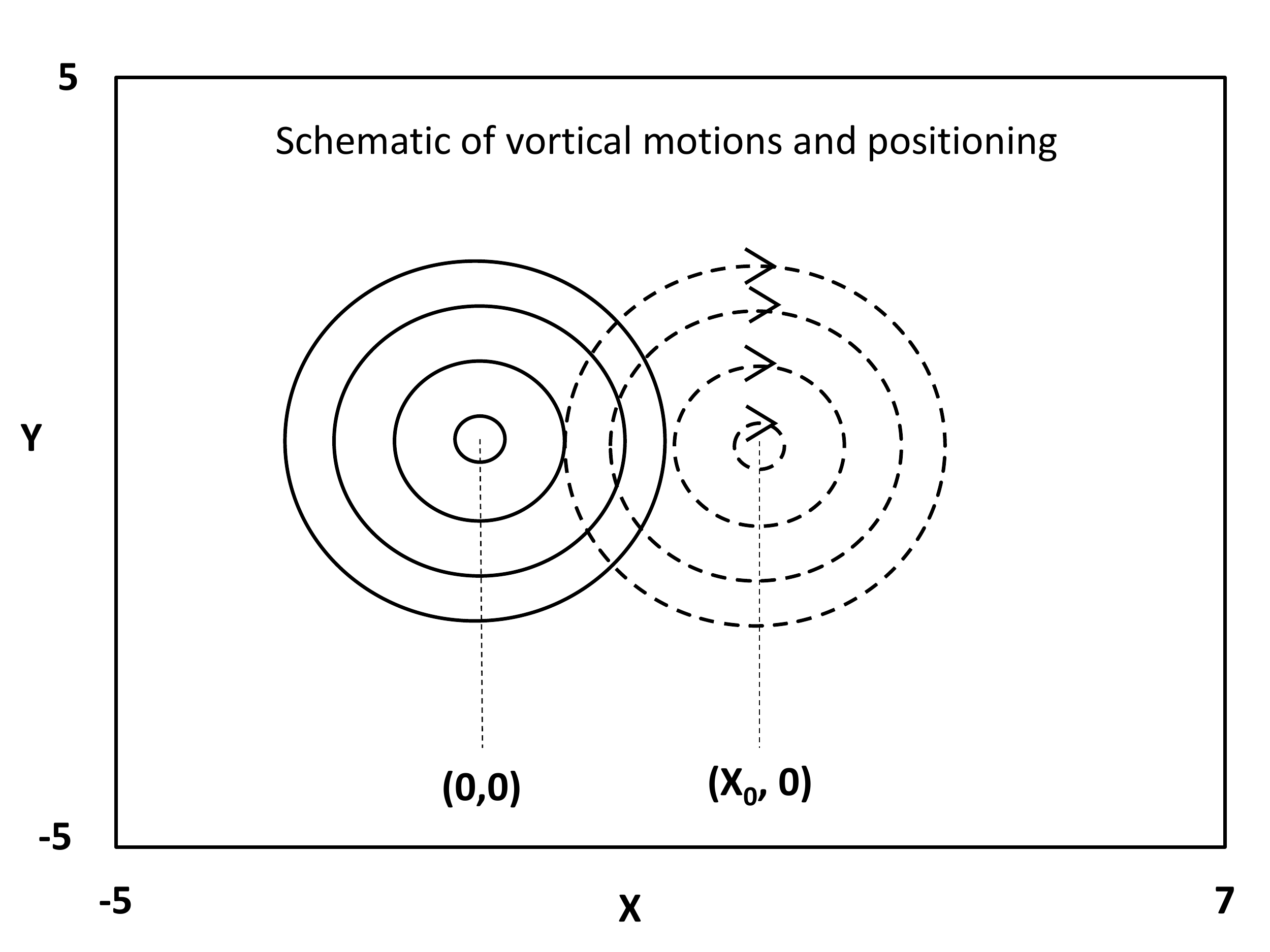}
\caption{Schematic illustration of boundary driving velocity. Two rotational motions are applied in sequence, first that illustrated in solid lines and then that in dashed lines. The direction of the solid line vortex in cases $1$A,B and C is anticlockwise and and clockwise in cases $2$ A, B and C. The position $(x_0,0)$ of the centre of the dashed line vortex is varied, but always twists clockwise. \label{basic_concept}}
\end{figure}

\begin{center}
\title{{\bf Table 1}: Driver Specifics} 
\begin{tabular}{cccccc}
\bf{Run} & \bf{Circulation} & \bf{Helicity}& \bf{$x_0$} & \bf{Entropy ($95\%$ C.I)} & \bf{Category}\\
\toprule
\bf{1A} & Opposite & Low & 0.5 & $0.161 \pm 0.004$ & Complex\\
\midrule
\bf{1B} & Opposite & Low & 1 & $0.786 \pm 0.017$ & Complex \\
\midrule
\bf{1C} & Opposite & Low & 2 &  $1.321 \pm 0.028$ & Complex \\
\midrule
\bf{2A} & Equal & High & 0.5  & $0.265 \pm 0.104$ & Coherent\\
\midrule
\bf{2B} & Equal & High & 1 & $0.444 \pm 0.046$ & Coherent\\
\midrule
\bf{2C} & Equal & High & 2 & $0.888 \pm 0.039$ & Coherent\\
\bottomrule
\end{tabular} 
\end{center}

One important feature to point out is the maximum speed attained by the driver. We wish to consider DC heating, therefore we choose to drive on a timescale slower than the Alfv\'en timescale. Specifics are chosen such that the maximum speed reached by the drivers about  is $0.1$, compared with an Alfv\'en speed of $1$. The period of the driver is around $75$ time units and the Alfv\'en travel time along the length of the loop is about $50$ time units.

\subsection{LARE3D}
We utilise the Lagrangian remap code LARE3D to evolve the system under the influence of the driver motions. A detailed description of the code is found in \citet{lare3dhard}. The code can be downloaded from http://ccpforge.cse.rl.ac.uk/gf/  along with a user guide. Here we briefly describe the main relevant features.\\
We solve for the resistive MHD equations in normalised form:

\begin{eqnarray}
\frac{D \rho}{D t} &=& -\rho \nabla \cdot {\bf v} \\
\frac{D {\bf v}}{Dt} &=& \frac{1}{\rho}(\nabla \times {\bf B}) \times {\bf B} - \frac{1}{ \rho}  \nabla P \\
\frac{D {\bf B}}{Dt} &=& ({\bf B}\cdot \nabla){\bf v} - {\bf B}(\nabla \cdot {\bf v}) \nonumber \\
                                 & - & \nabla \times (\eta \nabla \times {\bf B}) \\
\frac{D \epsilon}{D t} &=& -\frac{P}{\rho} \nabla \cdot {\bf v} + \frac{\eta}{\rho} j^2  \\
\nabla \cdot {\bf{B}} &=& 0
\end{eqnarray}
where $\bf{B}$ is the magnetic field, $\bf{v}$ is plasma velocity, $\rho$ is plasma density, $\eta$ is the resistivity and $j^2$ is the current density squared. The variable $P$ is the pressure which takes the form  $P = \rho \epsilon( \gamma - 1)$, where $\gamma$ is the ratio of specific heats,  and $\epsilon$ is the internal energy density.\\
At this stage we neglect gravity, conductive and radiative effects and stratification is ignored in order to save computational time in the long duration runs. Table $2$ contains specifics. \\
The full MHD equations are solved using finite differencing. The method involves setting up variables on a staggered Eulerian grid. Each timestep is divided in two: first the predictor computes an Eulerian estimation of the variables, and then the Lagrangian corrector updates these values, with second order accuracy in time and space. The last step is to remap the updated grid of information back on to the Eulerian grid.  A shock capturing viscosity is used such that in the event of shocks forming in the domain, the viscosity acts to smooth out the discontinuity in the localised region of the shock.\\

\begin{center}
\title{{\bf Table 2}: Simulation specifics common to all runs} 
\begin{tabular}{cc}
\bf{Property} & \bf{Value/Nature} \\
\toprule
x, y boundaries & periodic \\
\midrule
Upper z boundary & line-tied  \\
\midrule
Lower z boundary & driven, $\|{\bf v}\| \le 0.1 v_{A}$ \\
\midrule
$\bf{B}_0$ & $1 {\bf e_z}$\\
\midrule
$\bf{v}_0$ & $ \bf{0}$  \\
\midrule
Resolution & $256^3$ \\
\midrule
Domain & $[-5,7] \times [-5,5] \times [0,50]$ \\
\midrule
$\eta$ & 0.0005, uniform \\
\midrule
viscosity & shock capturing form \\
\midrule
Duration &  \parbox{5cm}{45000 time units (597 driver \\
  periods, 900 Alfv\'en times)} \\
\bottomrule
\end{tabular} 
\end{center}

As shown in the table each of the simulations is run for around $900$ Alfv\'en crossing times, which we have found to be a long enough time period to allow us to reach a statistically steady state. 
 
\section{Results}
On examination of the simulation results, we find that the low and high helicity cases exhibit very different behaviours. The low helicity case runs reach statistically steady states and we see steady heating of the loop, at levels corresponding to the ordering by topological entropy. On the other hand, the high helicity cases have less consistent behaviour, with evidence for intermittent instabilities and typically higher levels of heating. We describe each of the situations in turn in the following sections. 

\subsection{Low Helicity - Comparing runs \bf{1A,1B,1C}}
\begin{figure} [!h]
\includegraphics[width=\textwidth]{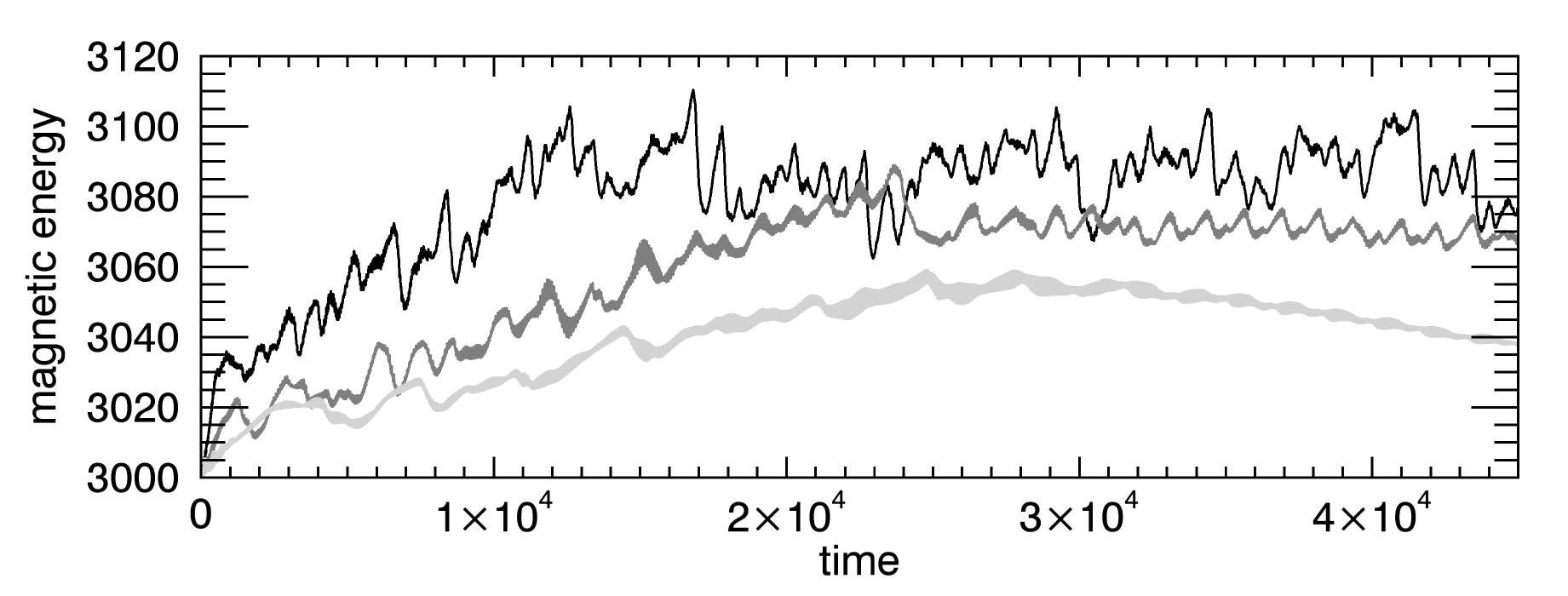}
\includegraphics[width=\textwidth]{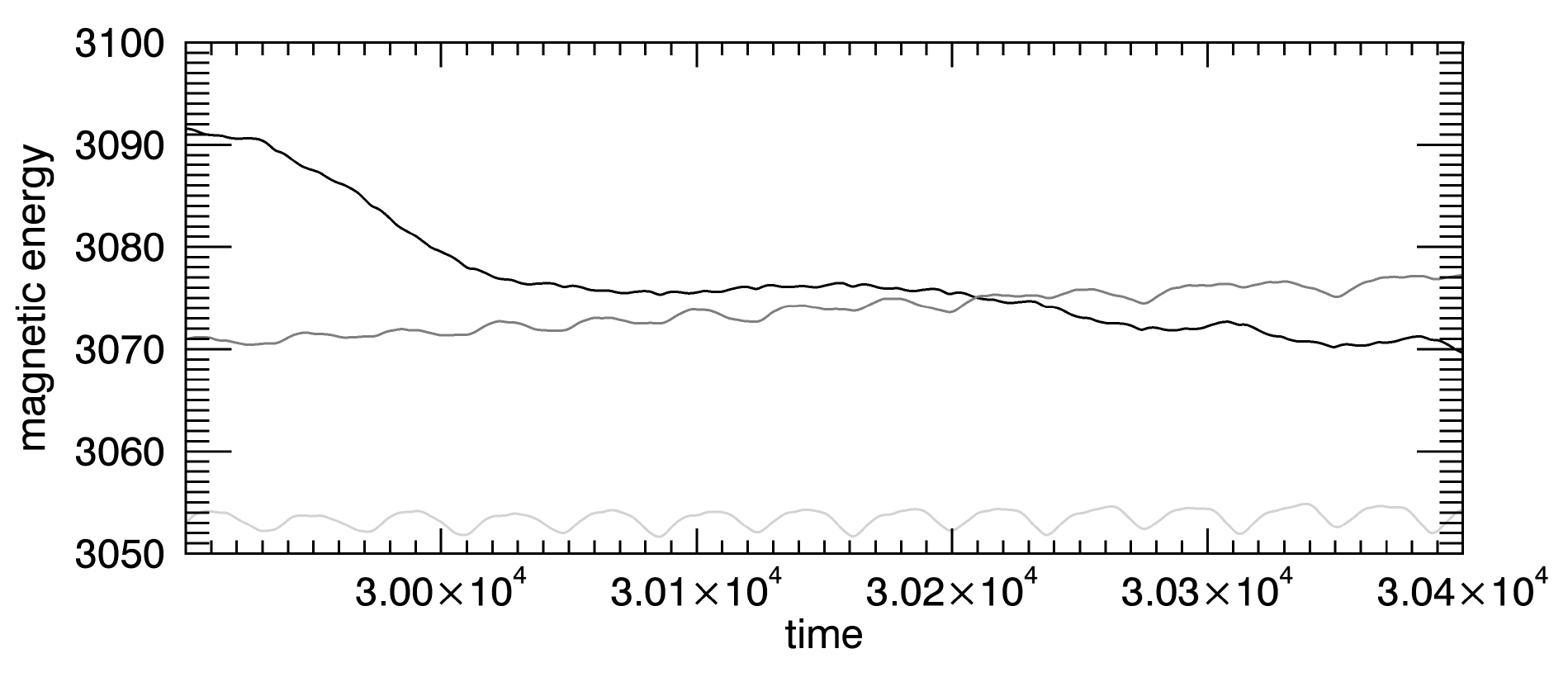}
\caption{Comparison of magnetic energy over the simulation for the three different values of $x_0$ in the low helicity cases 1A, B and C. The dark black line represents $1$C, the dark grey $1$B and the light grey 1A. It is clear that the most complex (1C) of the three cases receives the largest magnetic energy injection. We also observe the quantity reaching statistically steady states after about $20 000$ time units, with oscillations about an average thereafter. The steady states phase of the first graph is re-printed to show a section of the runs with the same line colours as before. The smaller scale oscillations in $1$C seem to correspond with the driver period.\label{enb_low_h} }
\end{figure}

Figure \ref{enb_low_h} compares the total magnetic energy evolution for our three different low helicity drivers (distinguished by different vortex centre positions $x_0$). The key feature to note in all cases is that the runs reach statistically steady states where the magnetic energy averaged over 100 driver cycles ($\approx$ 7500 time units) remains roughly constant. After about $20 000$ time units, or $400$ Alfv\'en times, during which we see an overall steady rise, the magnetic energy begins to display behaviour with a regular pattern. There is a bursty profile to all three lines, resulting from repeated small energy release events. Here and through the rest of the paper, where lines appear to be very thick due to the high number of oscillations in the line profiles we also plot an enlarged section of the run. From Figure \ref{enb_low_h}b we see that oscillations during lowest energy run, $1$A, occur roughly in correspondence with the twist of one vortex, i.e half the driver period (around $38$ time units). As we move to run $1$C however this correspondence is lost. \\
The next point to note is that the ordering of the average magnetic energy level with $x_0$  corresponds directly to the ordering of the topological entropy estimates. The position $x_0 = 2$ was shown to give the most complex driver profile, and it is here we see the largest values of magnetic energy. On the other hand, run $1$A for $x_0=0.5$ had the lowest entropy out of these three set-ups and displays the lowest values of magnetic energy. While all runs necessarily have magnetic energy in excess of the  potential 3000 units, we find only a low energy above potential builds up. Taking an average after $25000$ time units shows that run 1A with $x_0=0.5$ has magnetic energy about $1.6\%$ in excess of potential, while we have about $2.3 \%$ and $2.9\%$ in excess for 1B, $x_0=1$ and 1C, $x_0=2$, respectively. The higher topological entropy runs correspond directly to a magnetic field with higher energy.

We can compare the release of magnetic energy in the simulations with a real coronal loop plasma values. The potential field magnetic energy is given by $ B_0^2 V / 2\mu_0$
where $B_0$ is the field strength, $\mu_0$ is the usual permeability of free space and $V$ is the volume of the domain. Taking $B_0 = 100G = 10^{-2} T$, $V=10Mm \times 12Mm \times 50Mm$ and $\mu_0 = 1.2566 \times 10^{-6} H m^{-1} $ gives the magnetic energy in the potential field to be $2.387 \times 10^{23} J$. In the dimensionless units of the simulation the magnetic energy of the initial potential fields is $3000$, hence $1$ unit of magnetic energy in the simulation corresponds to $7.957 \times 10^{19}J$. If we look at the steady state section of run $1$C we see drops in the magnetic energy of around $10$ units. In a real coronal loop this would release $7.957 \times 10^{20}J$. This corresponds to the amount of energy released in a typical average flare.

\begin{figure} [!h]
\includegraphics[width=\textwidth]{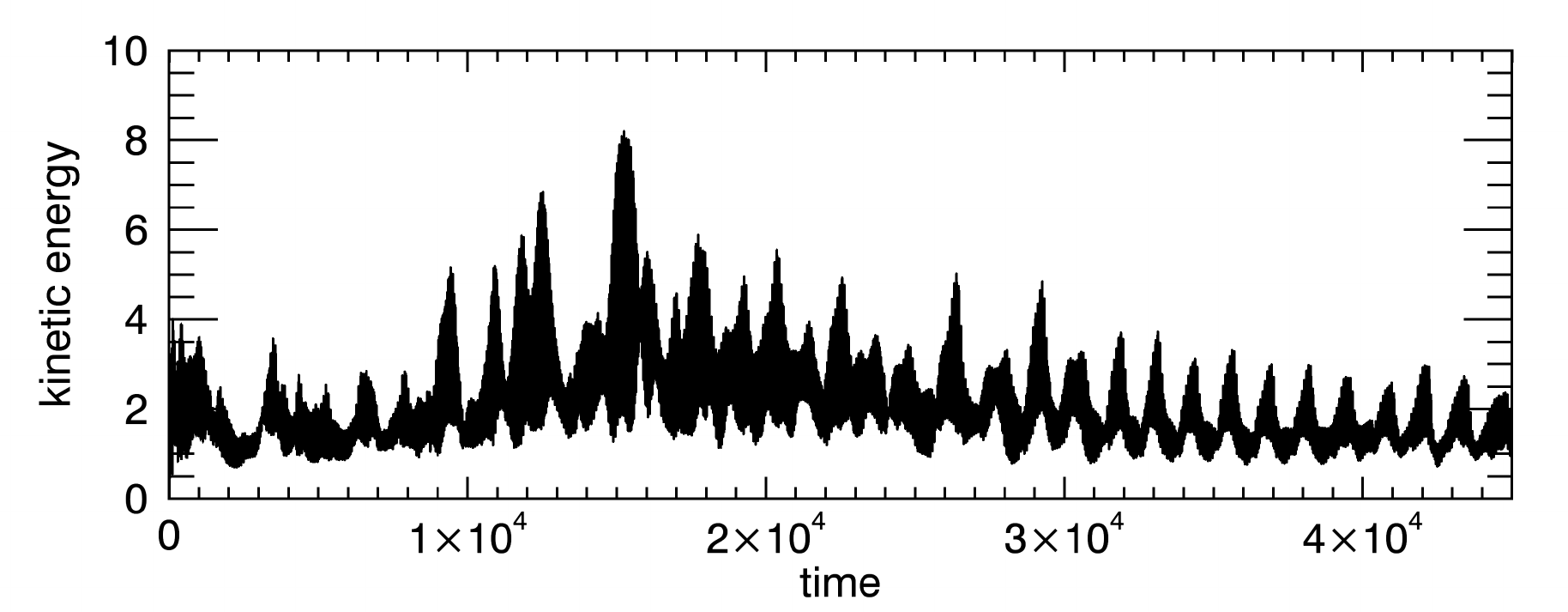}
\includegraphics[width=\textwidth]{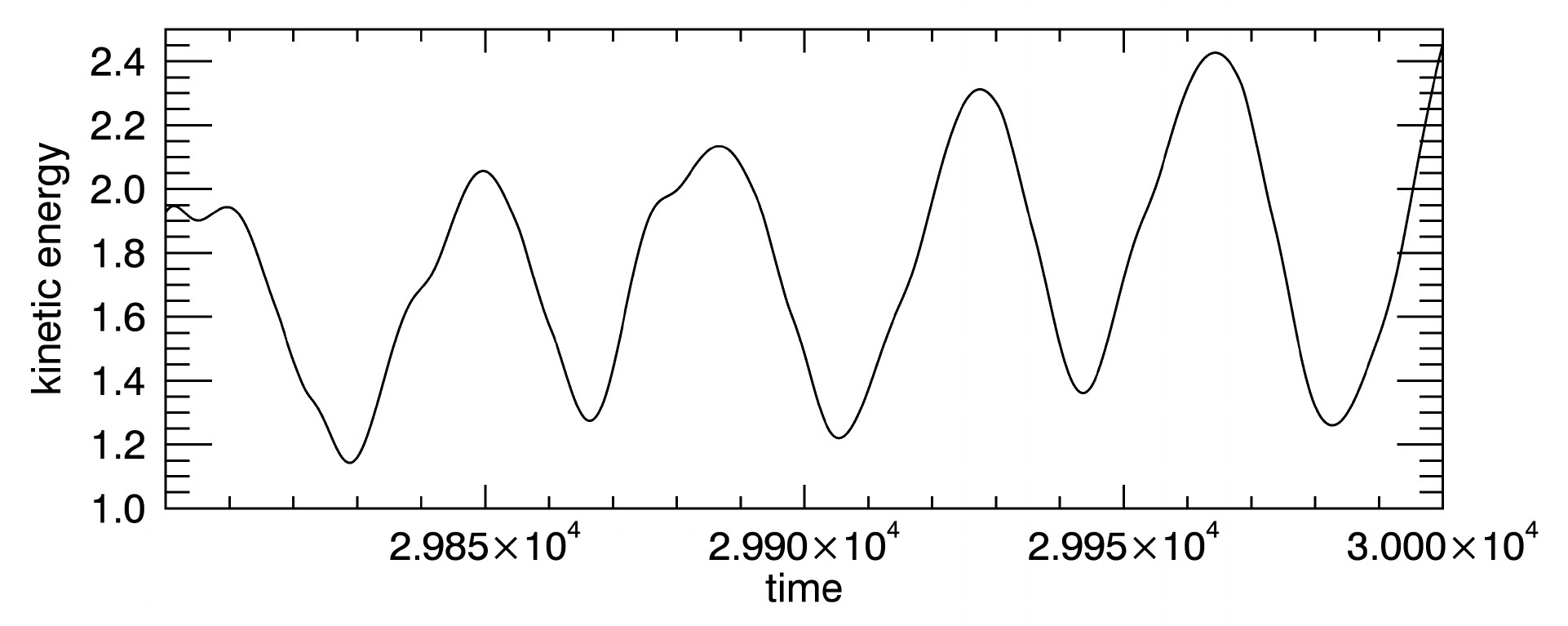}
\caption{Comparison of kinetic energy over the simulation for $x_0=1$ in the low helicity case. The values are much lower than for magnetic energy. On smaller time scales we observe peaks in kinetic energy corresponding to driving. \label{ke_low_h} }
\end{figure}

We now proceed to consider the kinetic energy evolution. Figure \ref{ke_low_h} shows the kinetic energy for run $1$B,  $x_0=1$. We see oscillatory behaviour, which at closer inspection is revealed to have time between peaks around roughly $38$ time units, corresponding to around half the driver period, or to one vortex spin. Comparing the magnitudes it is clear that the magnetic energy in excess of potential dominates over kinetic energy by a ratio of around 30:1 (taking an average of the magnetic and kinetic energy for $x_0=1$). These findings of statistically steady states with fluctuations and domination of magnetic energy over kinetic energy have previously been found as common features of continually driven systems. Similar behaviour in the kinetic energy has been found in papers such as \citet{Rapp2007} and \citet{Rapp2008}. However, the size of fluctuations in the magnetic and kinetic energy profiles do seem to correspond. Decreases in the magnetic energy occur around the same times as increases in the kinetic, suggesting that some magnetic energy is converted into kinetic energy. This additional kinetic energy may then be dissipated through viscous heating. \\
The new finding is the clear ordering by topological entropy of the mean magnetic energy. This raises the question of what exactly determines the mean level - future work could consider if there is some direct proportion with complexity.

\begin{figure} [!h]
\includegraphics[width=0.3\textwidth]{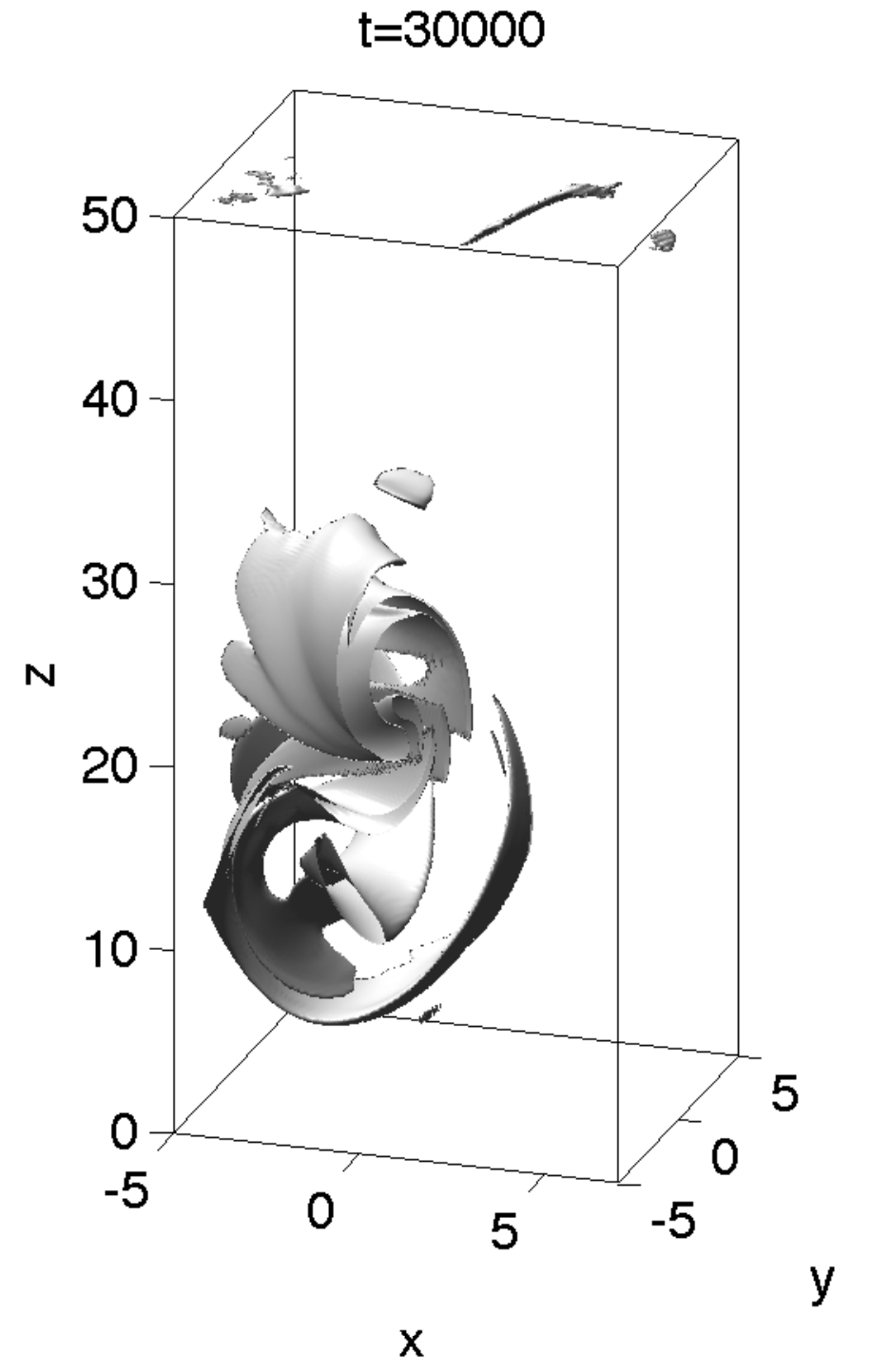}
\includegraphics[width=0.3\textwidth]{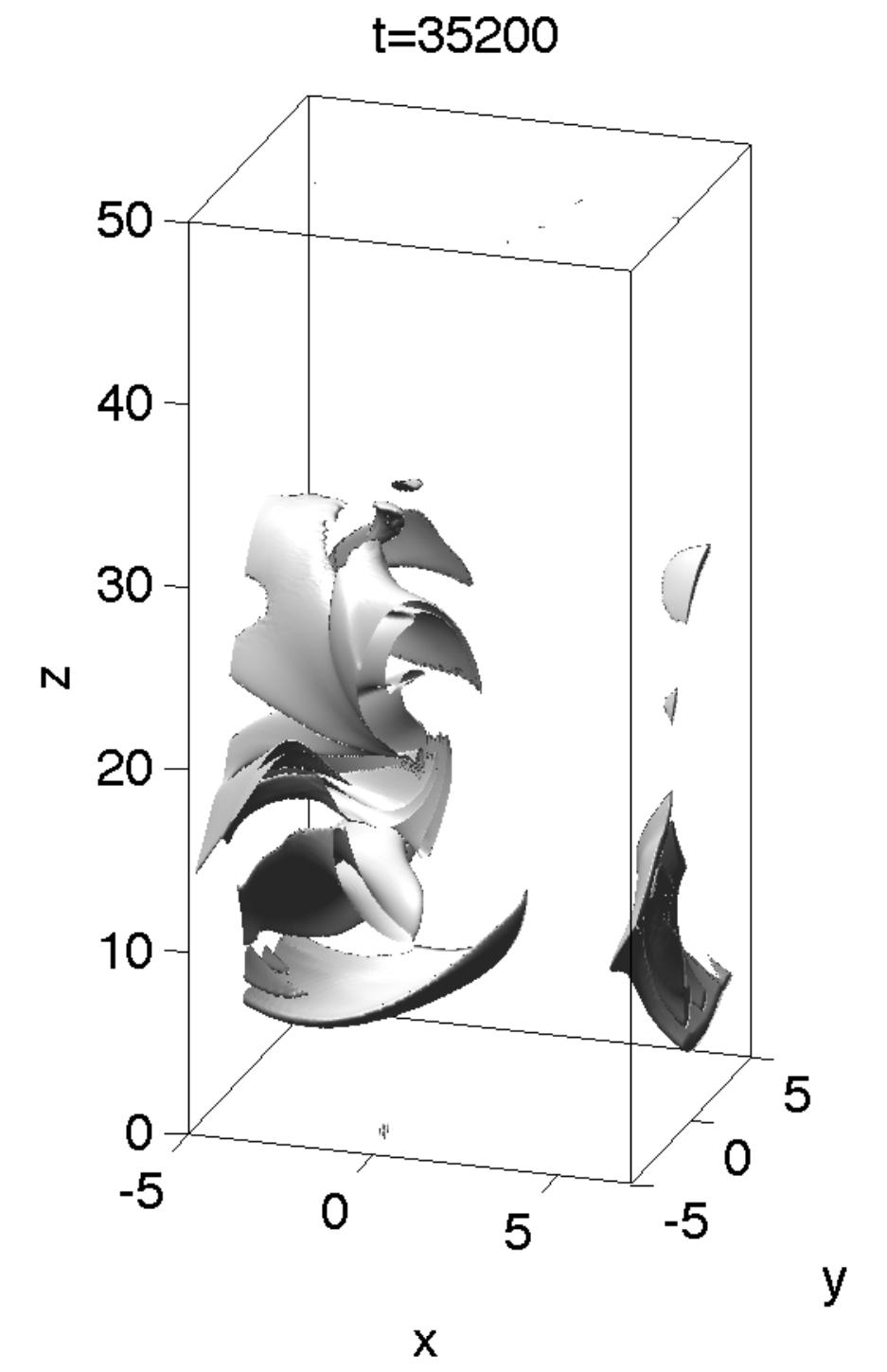}
\includegraphics[width=0.3\textwidth]{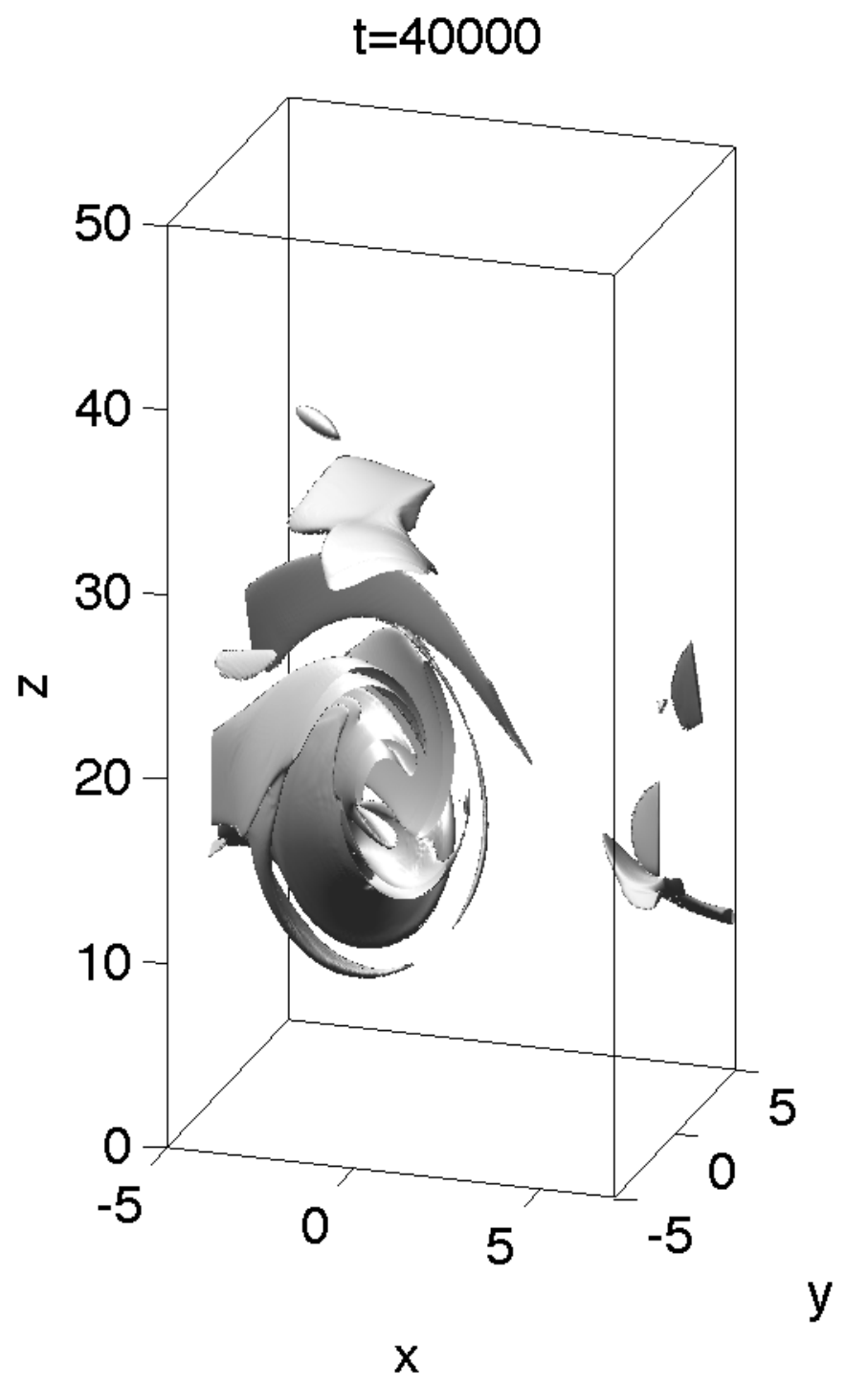}
\caption{Isosurfaces for $x_0=1$, complex case, at a threshold of current 0.4, around $20$ percent of the domain maximum of the first snapshot. \label{iso_x1opp} }
\end{figure}

The current structure of the field also exhibits new features. Figure \ref{iso_x1opp} shows for the run $1$B large, swirling structures, predominantly in the same regions. The isosurfaces are plotted for around $20\%~$ of the domain maximum of the first snapshot, which is a current of $0.4$. In Figure \ref{iso_x1opp} we see that the current is concentrated in thin, $3$ dimensional layers. The currents are generally in a certain region of the domain, spatially localised. This is similar to the findings of \citet{GalNord1996}. We see similar structure for all three times, which were taken from the statistically steady state section of the simulation. Typical values of the current density for this set of runs correspond to the ordering by topological entropy - stronger currents are generated in the higher topological entropy cases.  \\

 Many previous investigations using reduced MHD have found ribbons of current stretching between the upper and lower boundaries. These simulations have that $B_x,B_y << B_z$. Taking the $x_0=2$ run and examining the maximum values in the domain at four random different times gives that the maximum $B_x$ and $B_y$ range from $0.30-0.39$ and $0.28 - 0.35$ respectively, compared to a range over these same times for $B_z$ of $1.09 - 1.20$. So here we do not have $B_z$ significantly larger than $B_x$ and $B_y$. Therefore it appears that using fully $3$D MHD is important for obtaining the true spatial heating profile.

\begin{figure} [!h]
\includegraphics[width=\textwidth]{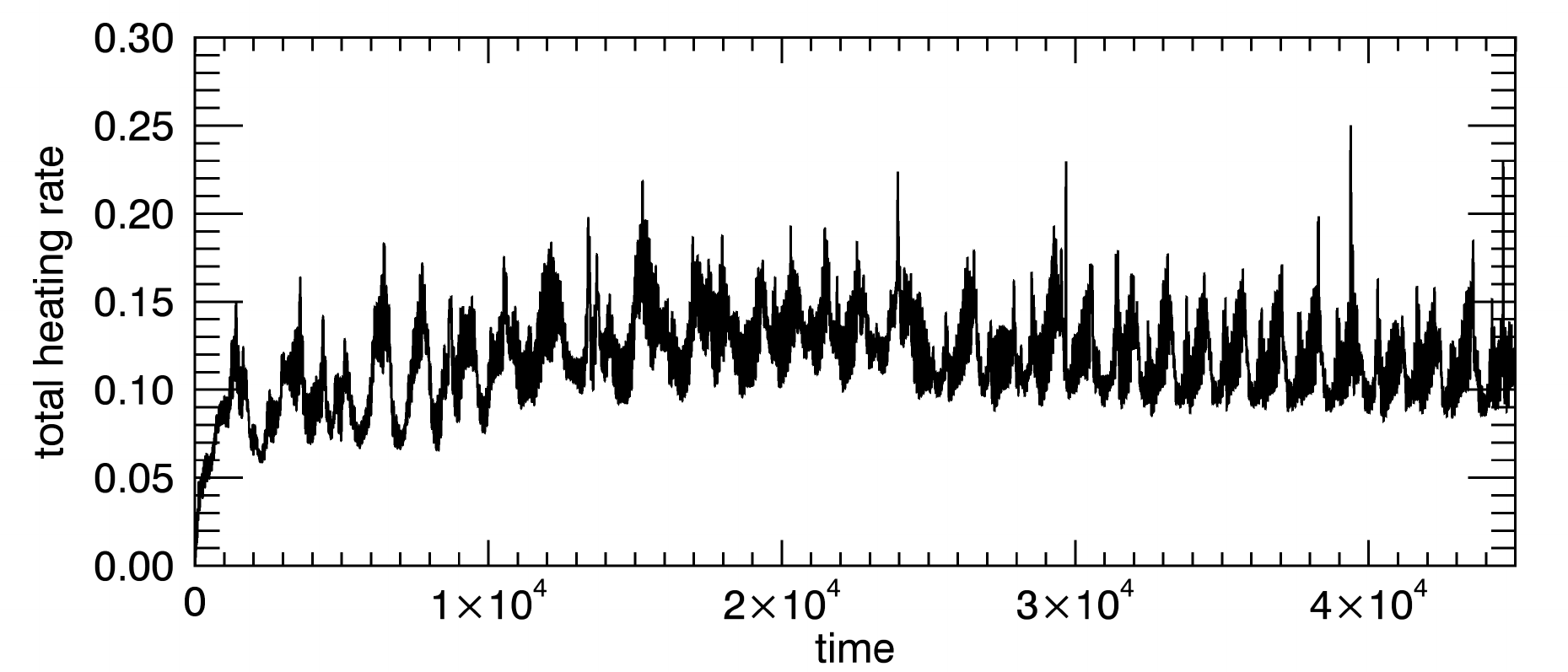}
\includegraphics[width=\textwidth]{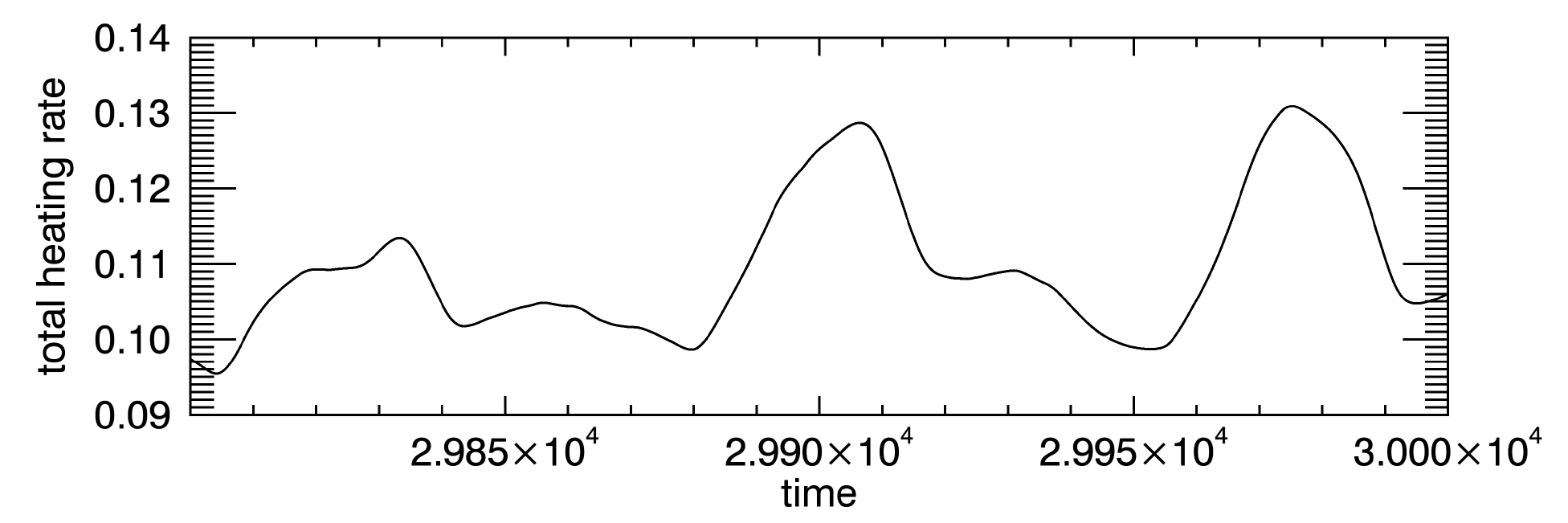}
\caption{Total heating rate for run $1$B. The heating takes the form of small but frequent bursts.  \label{low_h_heating}}
\end{figure}

We now turn to dissipation of energy in the loop. The ohmic heating does not consistently dominate over viscous heating, but we combine both the ohmic and viscous heating and consider a total heating of each run. It should be noted that the viscous heating is important, being a factor at shock sites, and therefore linked to reconnection and ohmic dissipation regions. For run $1$B, $x_0=1$, comparing the average viscous and ohmic heating gives a ratio of about $0.66$. For runs $1$A and $1$C the ratios are $1.12$ and $0.46$ respectively. Figure \ref{low_h_heating} shows the heating rate for the $x_0=1$ run. The mean values of the total heating rate are $0.18$ for the $x_0=2$ case, $0.12$ for the $x_0=1$ case and lastly $0.09$ for the $x_0=0.5$ case. The profile fluctuates around a fairly consistent average over the course of the simulation, with sections displaying regular behaviour. This low helicity, higher complexity configuration seems to result in a consistent pattern of low level heating. We find that the highest magnetic energy case is also producing the highest levels of heating. We build up more magnetic energy in the highest complexity case and that allows more energy to be intermittently but frequently released, giving the heating. Reconnection appears to kick in easily to unbraid the field slightly before the driving winds field lines up again. We should note that the time intervals between release points do not correspond to driver periods. \\

\subsection{High Helicity - Comparing runs \bf{2A,2B,2C}}

\begin{figure} [!h]
\includegraphics[width=\textwidth]{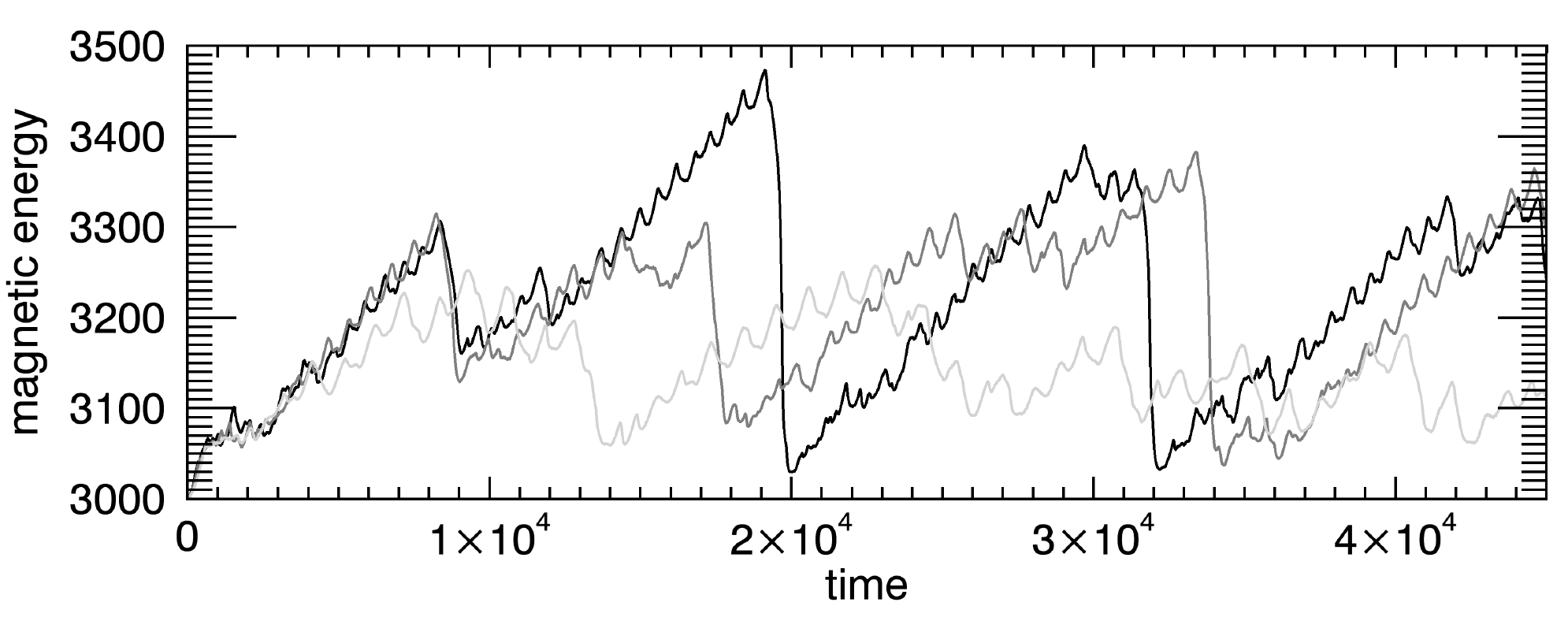}
\caption{Comparison of total magnetic energy in the high helicity cases 2A,B,C. The light grey line corresponds to $x_0=2$, the dark grey to $x_0=1$ and the black to $x_0=0.5$.  \label{enb_high_h} }
\end{figure}

Figure \ref{enb_high_h} illustrates the magnetic energy evolution for the high helicity cases. The main new result here is that a statistically steady state is not reached in these coherent cases. Instead the coherent motions are able to inject larger amounts of magnetic energy into the system. We see infrequent but very high energy release events at a few distinct times in the simulations for $x_0 =0.5$ and $x_0=1$, which we conjecture are times marking the onset of some global kink-like instability. In particular, papers \citet{Rapp2010} and \citet{Rapp2013}, using a reduced MHD system, found coherent motions, (shear and twist respectively) gave rise to statistically steady states after just one current driven instability (tearing, kink, respectively). These simulations ran for $600$ and then on the order of $1000$ Alfv\'en times respectively, so the models did run for relatively long times. Here we find instabilities occurring multiple distinct times, the first such demonstration in a continually driven system. \\
With the photospheric driver consisting of vortices centred at two distinct locations the magnetic field structure at any given time has no particular axis of symmetry. Hence the characteristic features of the well-known kink instability (see for example \citet{kink}, \citet{kinksim}) of twist exceeding a critical value and a subsequent kink of the central axis have no direct clear correspondence here. Nevertheless, we conjecture the coherent nature of the driver results in the input of a critical amount of twist following which a kink like instability takes place. A corresponding clear increase in kinetic energy is found, as would be expected in this circumstance, as discussed later in this section.\\
At the highest value of magnetic energy for $x_0=0.5$ we have an excess of $15.7\%$ in excess of the potential field. After this point we drop rapidly back to much nearer the potential of 3000 units. This behaviour is in complete contrast to all previous examinations of continuously driven systems, where statistically steady states are reached.\\
 While the primary feature of these cases is the big releases, secondary to this are regular smaller energy release events in the run-up to these large discharges of magnetic energy. This is similar to the complex case. Again note no direct correspondence with the driver period. Over the course of the largest change in magnetic energy, for run $2$A ($x_0=0.5$), the magnitude changes from $3473$ units to $3029$ units (i.e. about $93$\% release of the available free energy) over a time interval of $910$ units (corresponding to $18.5$ Alfv\'en times or 12 driver periods). \\
 In the case of $x_0=2$ however we only see the smaller, bursty events. To explain this consider the conceptual differences of this run. If $x_0=0$ then we would be in the situation where the field structure is twisted most coherently. If we were to take $x_0$ to infinity, we would have two systems in which two kinks would appear in twice the time. In our simulations $x_0$ varies between these two limiting cases, and the field is twisting less tightly. Our lowest entropy runs, $x_0=0.5$ and $x_0=1$, are closer to the first limiting case: they have low topological entropy and exhibit kink like behaviour. The highest complexity case, $x_0=2$, twists the field the least coherently out of the three runs. It is not close enough to either limiting case for the twisting to reach the critical state. 

Using the same arguments as before we can estimate what these simulation quantities would actually correspond to realistically. Say one of the larger events in run $2$C releases roughly $300$ units. This would translate into a flare of energy $2.387 \times 10^{22} J$, a larger than average event. 

\begin{figure} [!h]
\includegraphics[width=\textwidth]{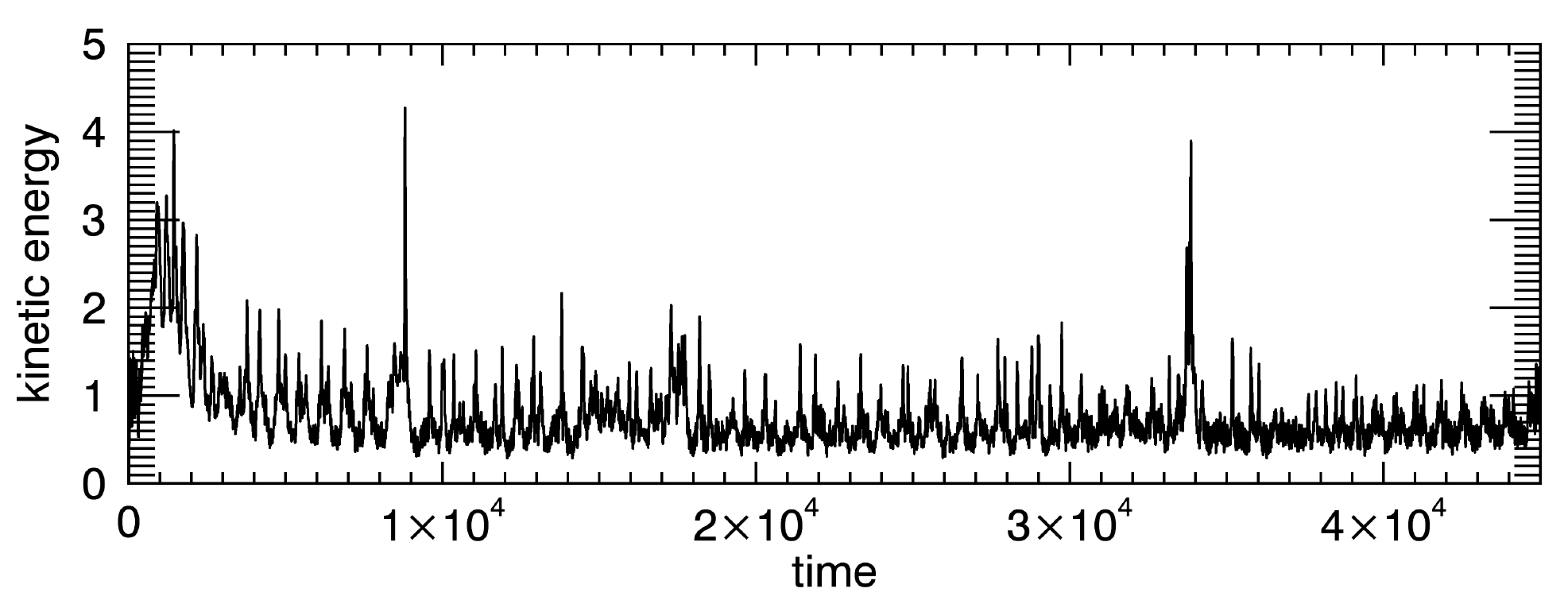}
\includegraphics[width=\textwidth]{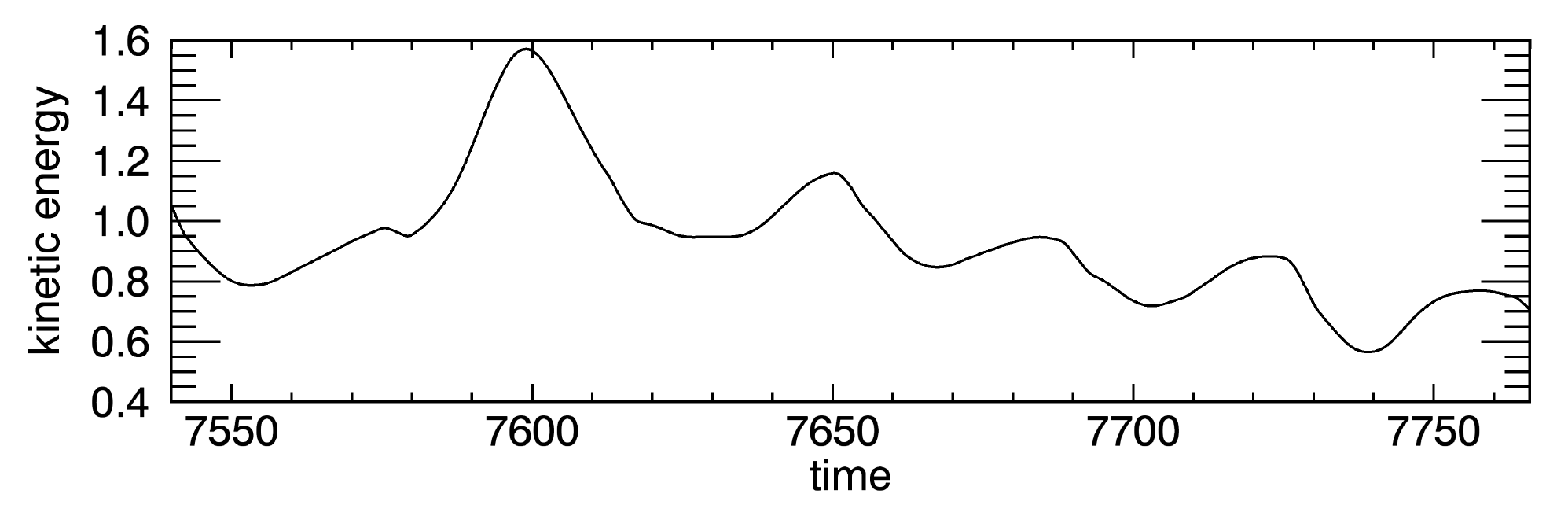}
\caption{Comparison of kinetic energy in the high helicity case 2B, $x_0=1$. \label{ke_high_h}  }
\end{figure}

We now turn to the kinetic energy. The kinetic energy for $x_0=1$ illustrated in Figure \ref{ke_high_h} shows clear, short duration bursts, as expected for the high magnetic energy release events resulting from the kink-like instability. The peak in kinetic energy corresponding to the largest drop in magnetic energy spans a time of around $120$ time units. This is $1.59$ driver periods or $2.4$ Alfv\'en times. However little of the magnetic energy is being released as kinetic energy - unlike before in the complex case the fluctuation sizes in magnetic and kinetic energy do not correspond. This points towards the idea that here more magnetic energy is being dissipated directly by ohmic heating, with less being converted to kinetic before viscous heating takes place. The kinetic energy shows smaller fluctuations about a lower level than in the complex cases, punctuated by the larger bursts at the same times as the magnetic energy drops in the $x_0=0.5$ and $x_0=1$ cases. Again free magnetic energy dominates over kinetic energy, the ratio now being time dependent with the distinct energy injection and energy release phases. 

\begin{figure} [!h]
\includegraphics[width=0.3\textwidth]{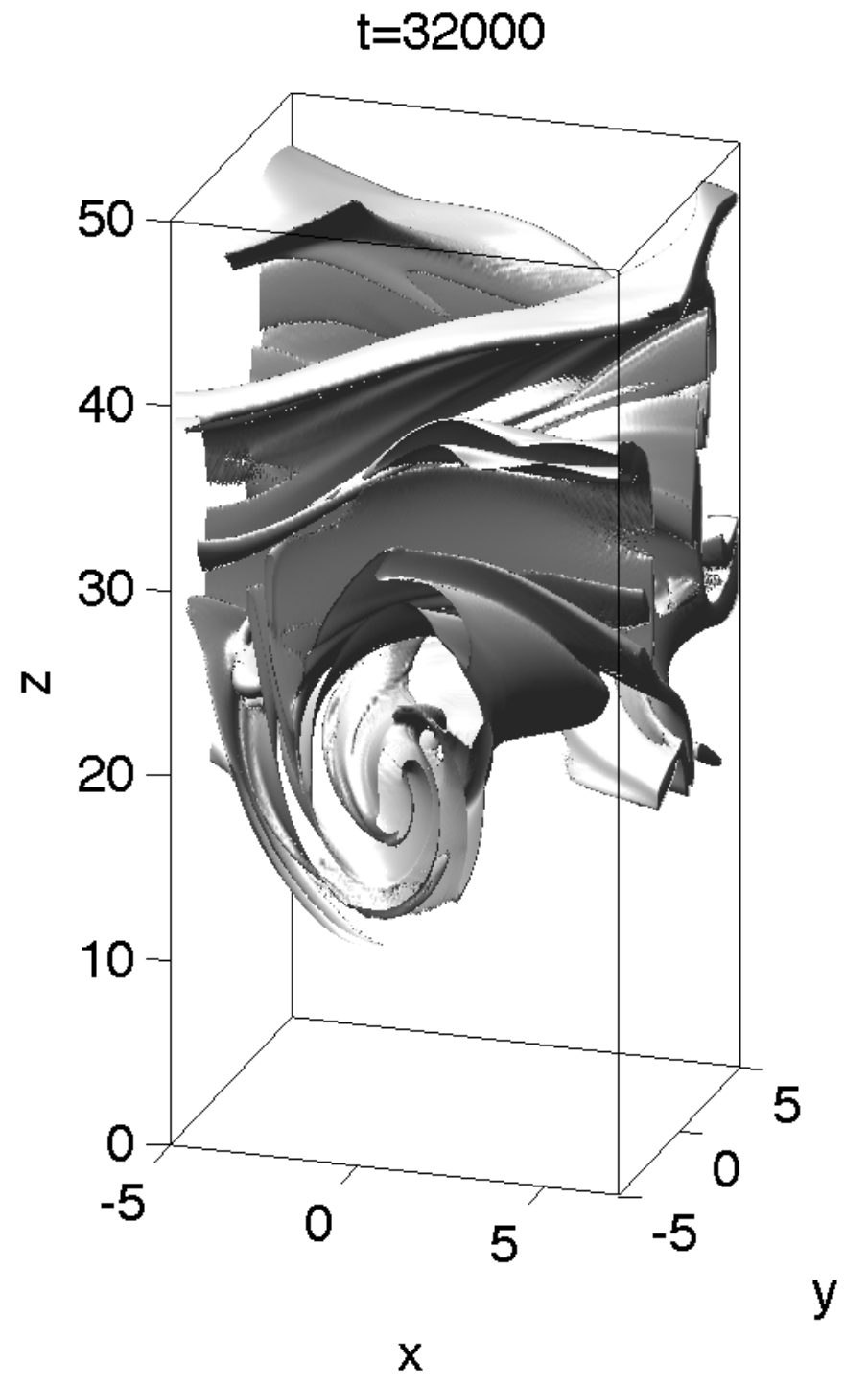}
\includegraphics[width=0.3\textwidth]{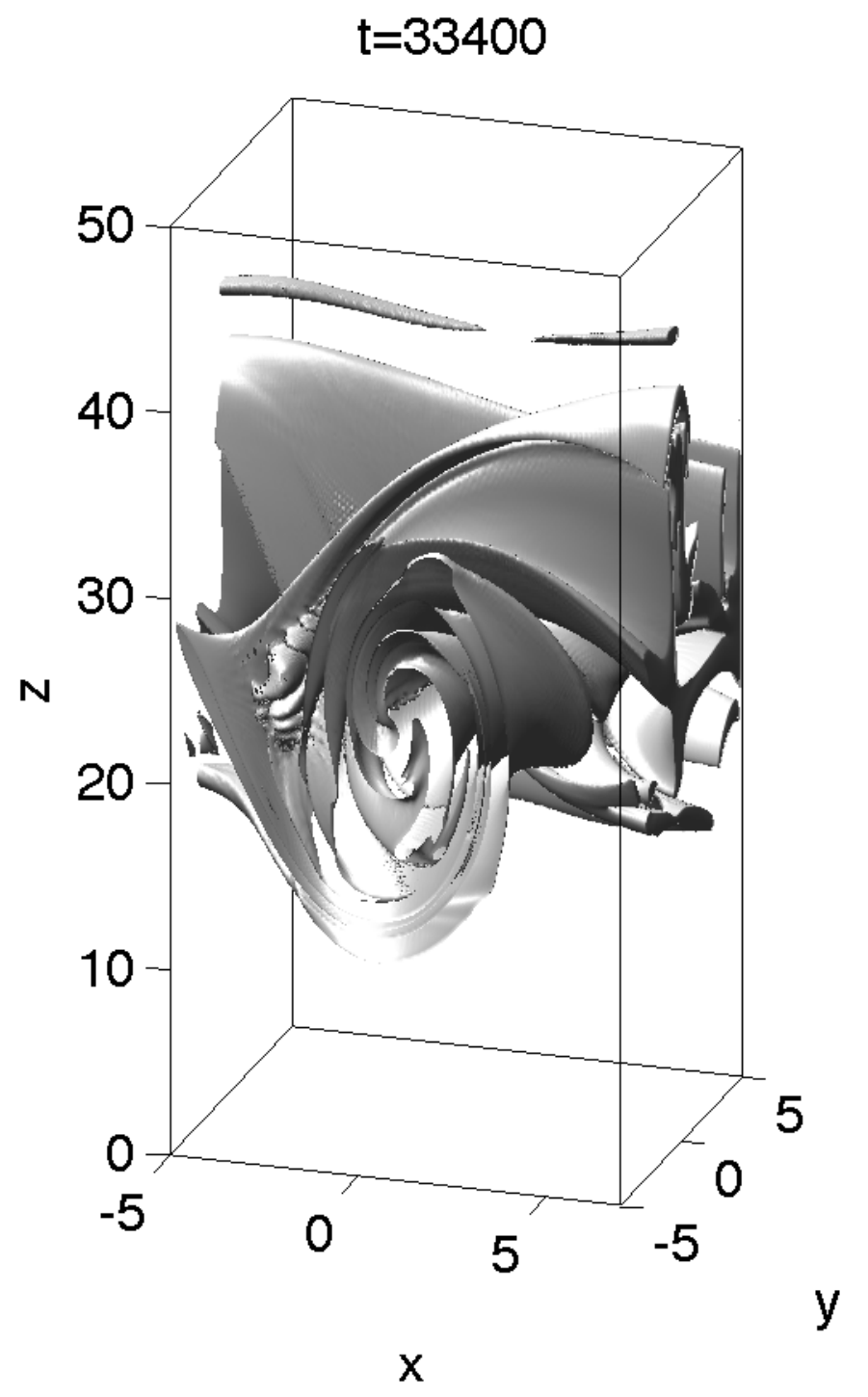}
\includegraphics[width=0.3\textwidth]{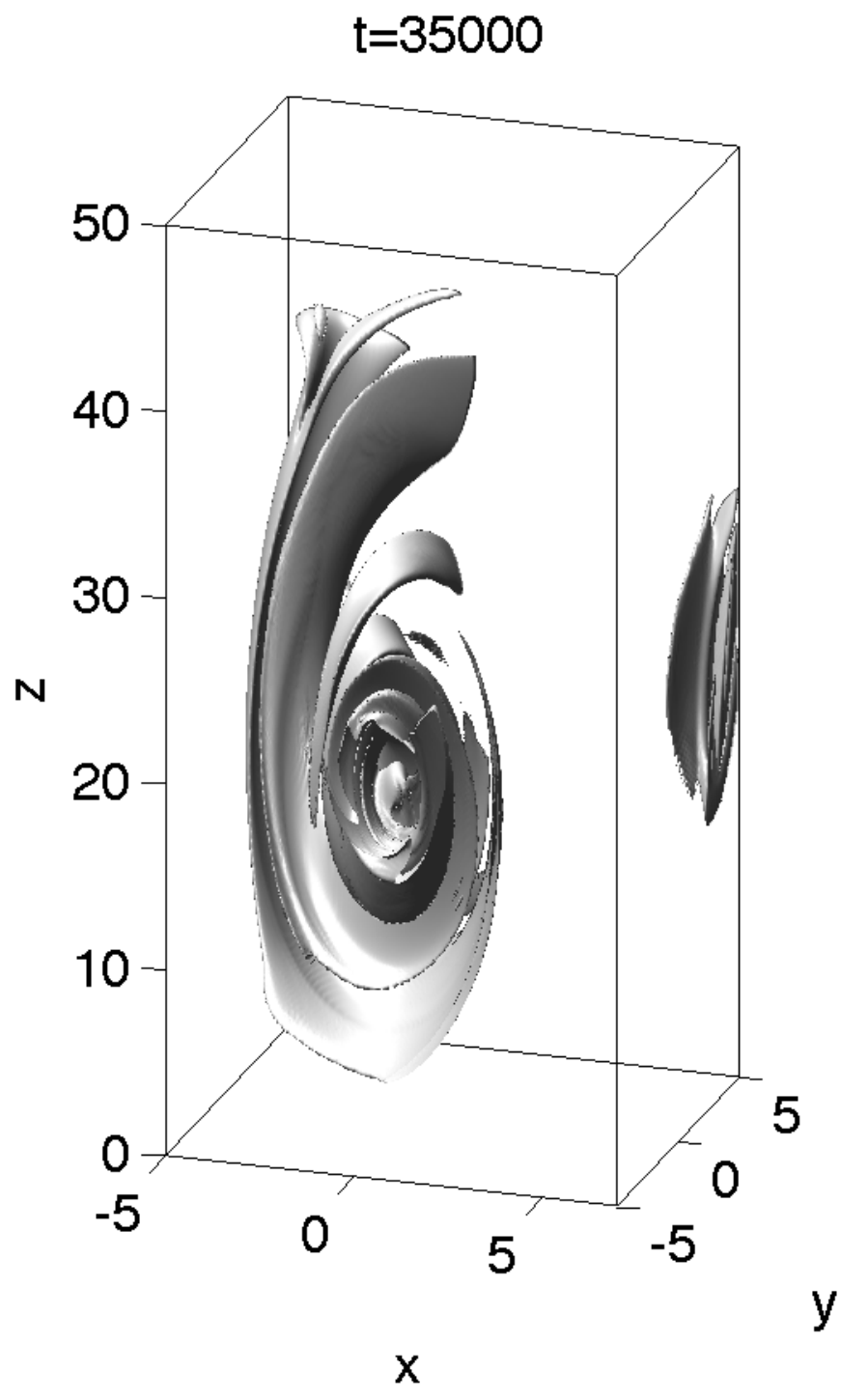}
\caption{Isosurfaces for $x_0=1$ for the high helicity case at low threshold, current value 0.4 \label{iso_x1eq} }
\end{figure}

\begin{figure} [!h]
\includegraphics[width=0.3\textwidth]{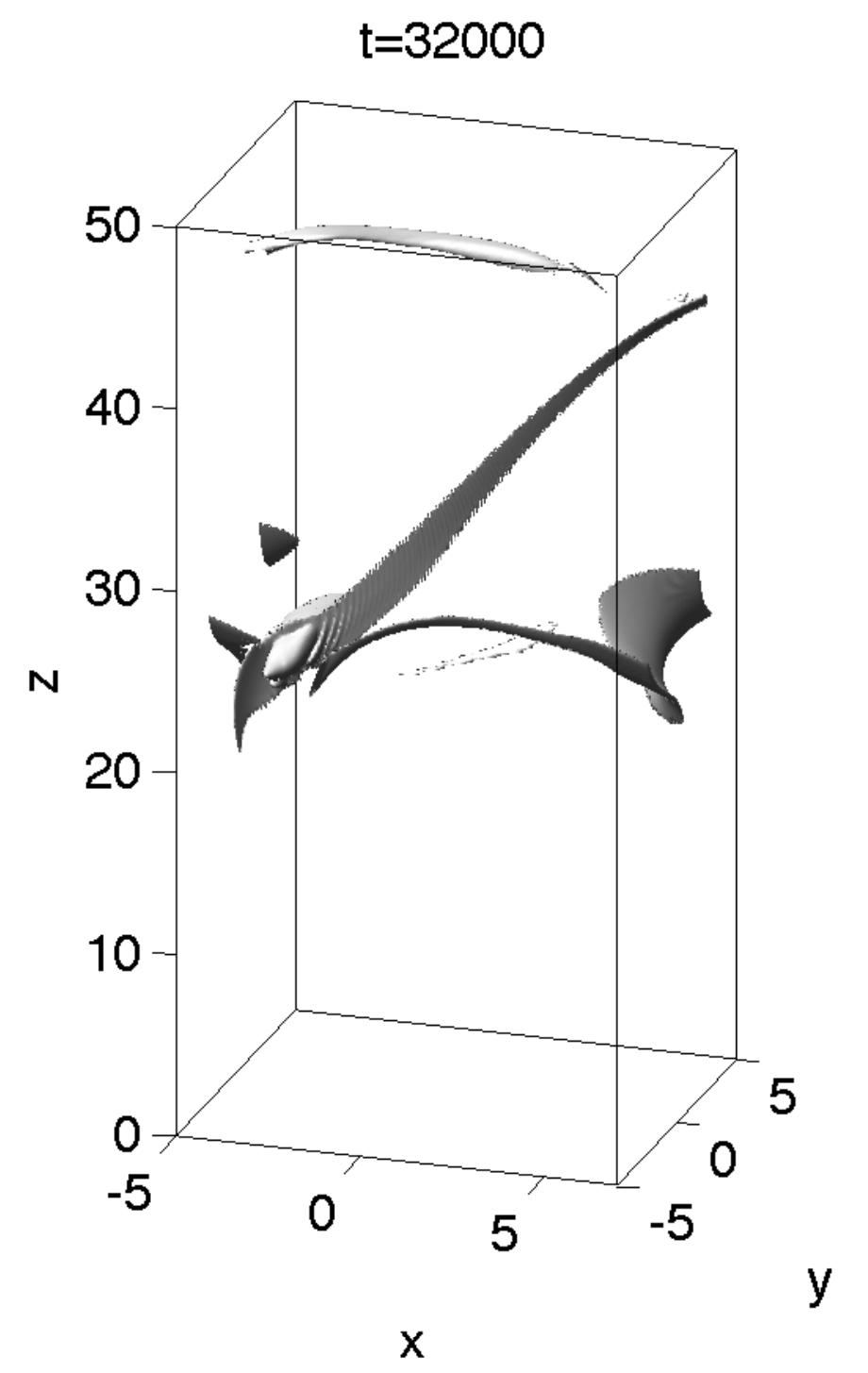}
\includegraphics[width=0.3\textwidth]{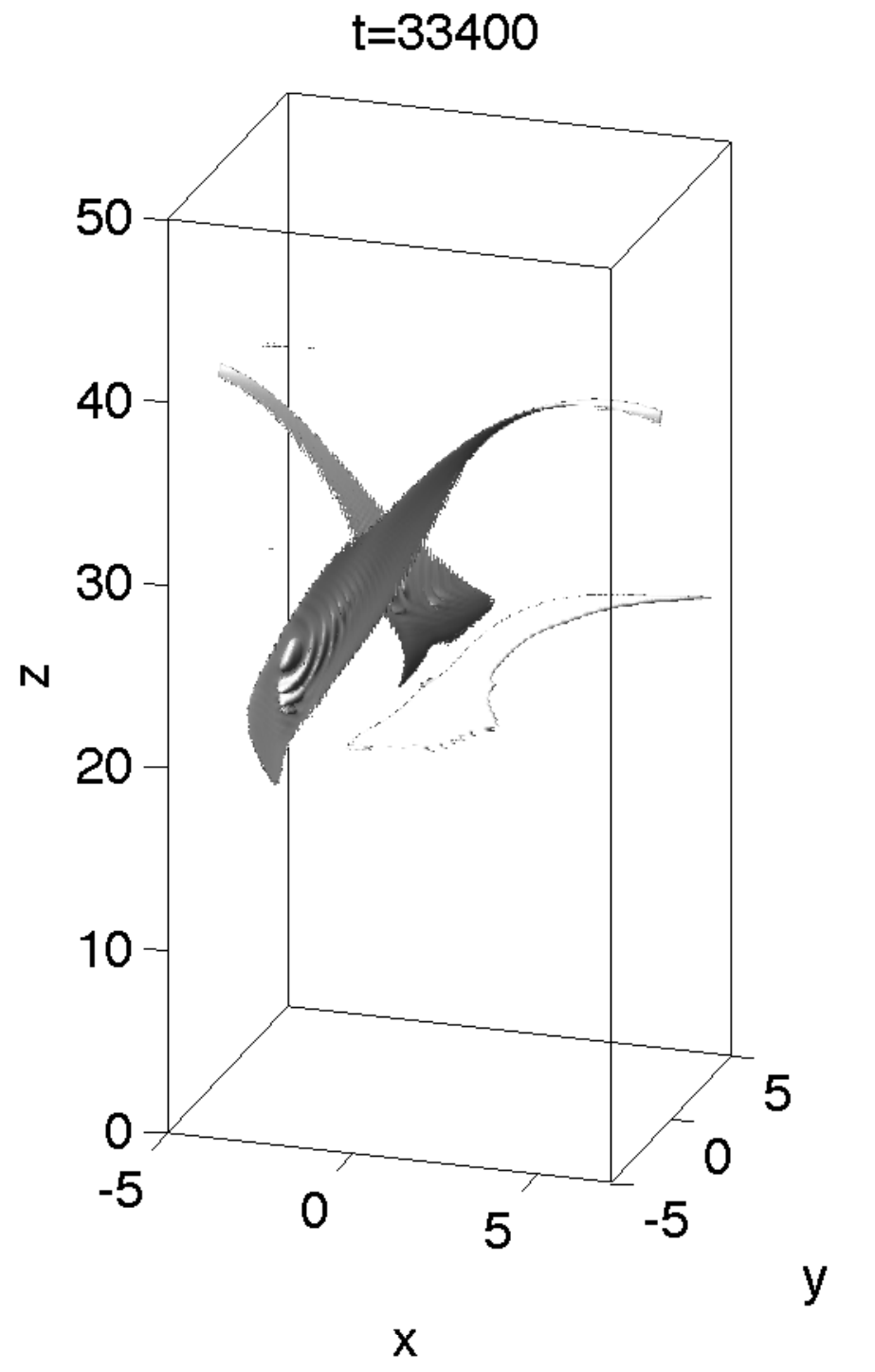}
\includegraphics[width=0.3\textwidth]{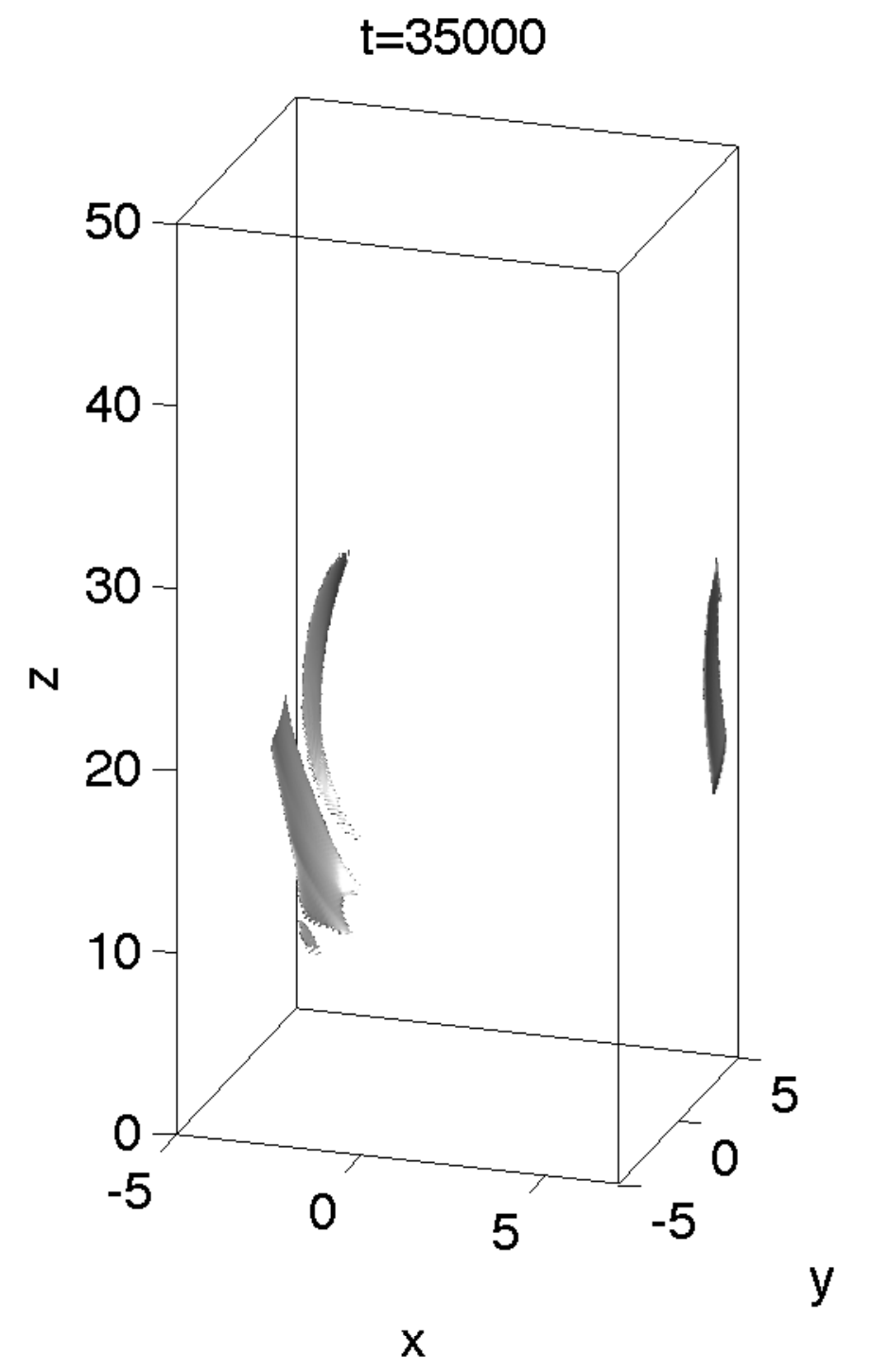}
\caption{Isosurfaces for $x_0=1$ for the high helicity case at higher threshold, current value 2. \label{iso_x1eq_ht} }
\end{figure}

Consider now the currents in the domain. Again we examine the structure by plotting isosurfaces at different times. Figure \ref{iso_x1eq} shows currents at times around the largest of the magnetic energy releases for the run $2$B, $x_0=1$. The isosurface is plotted at the same value as before, $0.4$. The first of the plots is before the release, so the field lines have been twisted to a large extent but not to the point of the instability. We see more volume filling by the current density, with swirled structures and ribbons running horizontally through the domain. Next we have a plot at a time as the large scale energy release is occurring. The volume filled by current at this level has lessened, and by the time we look at a point after the energy has reached the minimum of the event we have a much simplified, elongated swirl of current. We also consider current at a higher threshold in Figure \ref{iso_x1eq_ht}. The maximum current at these three times varies between $4.6$ and $21.9$, and we plot in all three cases the current isosurface at a value of $2$. Now we see much smaller fragments, indicating that the higher we take the threshold in this case, the more we observe small current layers, unlike for the low helicity case.

\begin{figure} [!h]
\includegraphics[width=\textwidth]{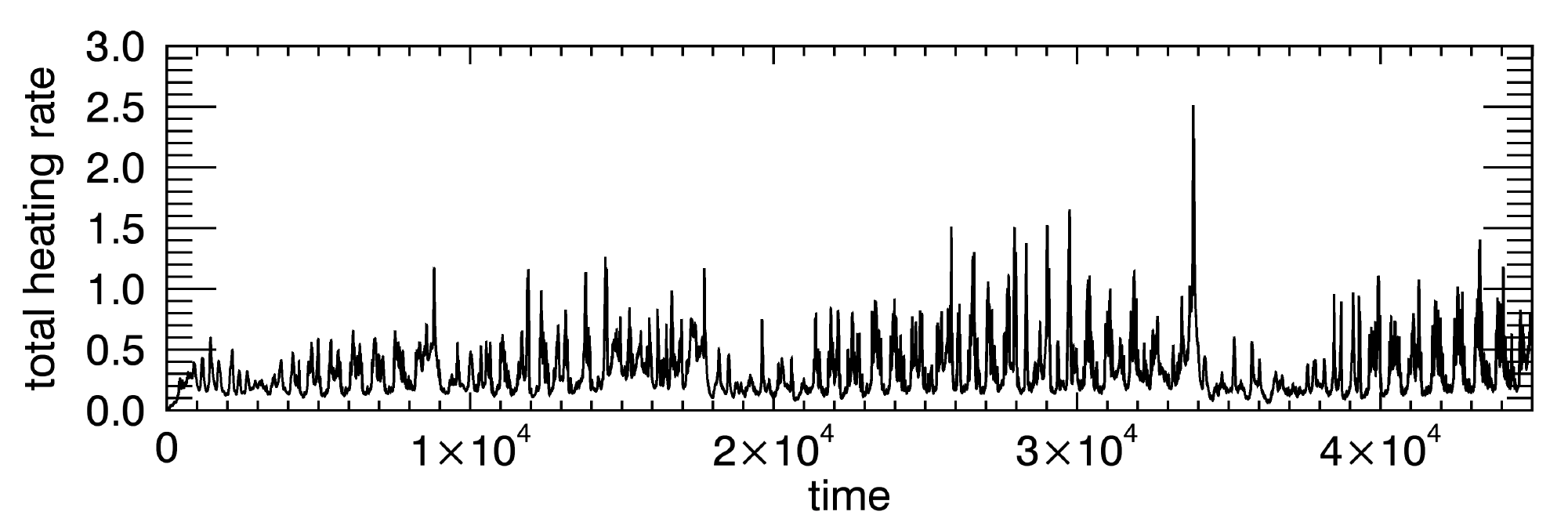}
\includegraphics[width=\textwidth]{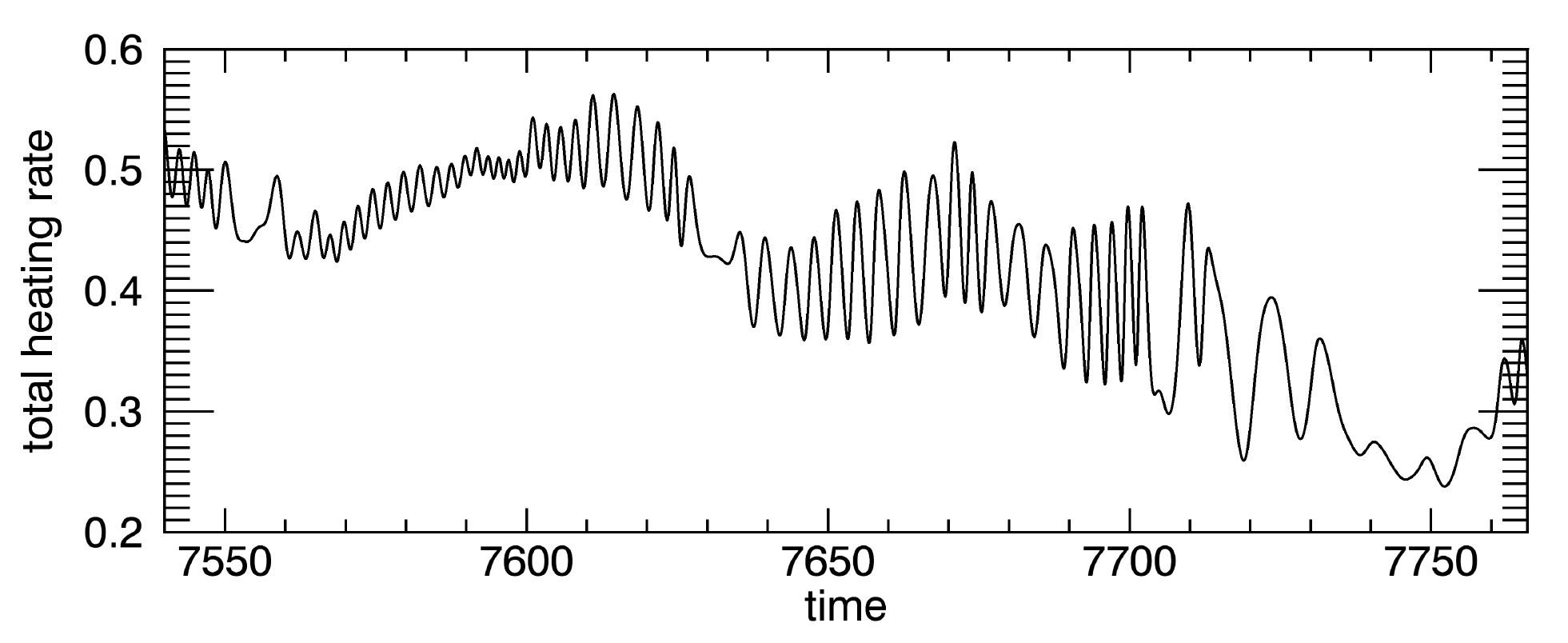}
\caption{Total heating rate for run $2$B, $x_0=1$. \label{high_h_heating}}
\end{figure}

Regarding the heating rate, Figure \ref{high_h_heating} shows the total heating rate for the $x_0=1$ case. Again we consider ohmic plus viscous heating, but note that this time the ratios of average viscous to average ohmic are $0.3$, $0.28$ and $0.30$ for run $2$A, $2$B and $2$C respectively - here viscous heating seems to increase in proportion with the ohmic dissipation. We see a profile with bursty characteristics, the largest spikes corresponding to the largest peaks in magnetic energy. The heating rate is much more variable and intermittent than for the previous set of runs. The average of the total heating rate for $x_0=0.5$ is $0.342$, with a maximum of $3.184$, larger than any previous values. The runs with $x_0=1$ and $x_0=2$ have average heating of $0.332$ and $0.235$ and maximum of $2.513$ and $2.406$, respectively, illustrating further that higher levels of helicity contribute to more heating even in a lower complexity state. At this point it would be inappropriate to consider temperature information since we have neglected conduction and radiation, however this would be of interest in future studies. 

\section{Discussion and Conclusions}
The aim of this work has been to understand the ways in which different types of photospheric motions that drive field line braiding can affect the nature of heating of coronal loops. We designed a series of drivers with varying complexity and ability to inject helicity into a domain. These consisted of two blinking vortices lying in the xy plane. They fell into two main categories - high helicity (coherent action by drivers) and low helicity (complex action by drivers). Within these two categories we varied the position of second vortex in order to create further incarnations of the flows. The low helicity drivers had typically higher complexities as measured by their topological entropy than their high helicity equivalents.\\
Long duration fully $3$D MHD simulations of a straightened coronal loop under the influence of these drivers showed some new and surprising features. It was found that the low helicity motions gave rise to a statistically steady state in which a low (relative to this work) level of heating was reached. Depending on the time scale of the driver this heating output could nevertheless be sufficient  to heat the corona. Additionally, the magnitude of magnetic energy and heating injected into the system corresponded directly to the measured topological entropy. Magnetic energy and heating profiles were ordered in terms of magnitude by $x_0=0.5$, $x_0=1$ and $x_0=2$, with entropy measurements increasing respectively. Here complexity seems to have been the dominant factor in the level of heating.\\
In our high helicity case we trigger a kink-like instability in our most coherent cases and do not settle to a statistically steady state. Furthermore, we do not see the same correspondence with entropy as before. The lowest entropy run, for $x_0=0.5$, has the highest magnetic energy and heating profiles. This is the most coherent case - the vortices are closest together so that the structure of the field is being twisted to higher degree, injecting the most Poynting flux. This situation resulted in steep increases in magnetic energy until conditions were reached to trigger an instability, releasing relatively large amounts of energy in the form of heat over just a few driver periods. The second most coherent is the run $x_0=1$. We see similar large discharges of magnetic energy but not on the same scale as before. Finally, our least coherent of these three runs, $x_0=2$, which was also the most complex of the three, does not display such large scale features. It would appear that here it is the helicity that is the key factor. \\
Coherent, low complexity but high helicity-injecting motions on the photospheric appear to be the most effective at heating coronal loops. Both sets of runs exhibit bursty magnetic energy profiles, showing that in all cases we have small reconnection events occurring at regular intervals. However it seems only the most coherent cases are able to twist field lines to a high enough degree to trigger some larger scale instability and supply higher levels of thermal energy. We have shown that the type of driver is in fact crucial in determining the nature of heating in a loop. \\
\underline{Acknowledgements}; The authors would like to acknowledge financial support from STFC, (grant code ST/K502443/1 and ST/K000993). Computations were carried out on the UKMHD consortium cluster funded by STFC and SRIF. We are also grateful for advice on topological entropy from Jean-Luc Thiffeault.

\section{Appendix 1 - Driver Specifics}
Here we present the full expression of the driver:

\begin{eqnarray}
v_x=& ky f (t)\exp((-x^2-y^2)/2) +& \\ \nonumber
         &yp(t)\exp((-(x-x_0)^2-y^2)/2) &\\ \nonumber
v_y =& -kxf(t)\exp((-x^2-y^2)/2)+ & \\ \nonumber
          &-(x-x_0)p(t)\exp((-(x-x_0)^2-y^2)/2)&
\end{eqnarray}
where $k=-1$ in the low helicity case and $k=1$ in the high helicity case, and

\begin{equation*}
f(t) = \begin{cases}
  -0.075\cos{t} + 0.075  & 0<t <\pi \\
      0.15                           &  \pi<t <11\pi  \\
       -0.075\cos{t} + 0.075  & 11\pi<t<12\pi  \\
       0                                &12\pi<t<24\pi 
\end{cases}
\end{equation*}

\begin{equation}
p(t)= \begin{cases}
  0 & t<12\pi \\
   -0.075\cos{t}  +0.075              &            12\pi<t <13\pi  \\
     0.1 5  & 13\pi<t<23\pi \\
       -0.075\cos{t} +0.075              &               23\pi<t<24\pi  
\end{cases}
\end{equation}
The driving velocity was extended to be infinitely time periodic.\\
The driving speed and duration were chosen to satisfy two criteria. The first was to give a maximum velocity around $0.1$ of the Alfv\'en speed, corresponding to slow driving. The second criterion was due to the fact that this work follows on from that in papers such as \citet{wsetal2011}. The twists in the braids examined had magnitude $\pi$, and so we chose parameters which advected a particle starting at the origin roughly minus $\pi$ with a twist of the first vortex.  \\
In principal however the rate of ramp up, period of maximum driving, amplitude of maximum driving and rate of ramp down can all be adjusted for the user's purposes. 
 At a time of maximum driving, the driver strength declines exponentially from a maximum of about $0.1$ at a radius of $1$ (reaching a value of $0.005$  at a radius of $r=3$).

\section{Appendix 2 - Topological Entropy}

Topological entropy is a measure of the chaos of a $2$D flow. It has many equivalent descriptions, but \citet{Newhousedefn} quotes \lq asymptotic growth rate of material lines' . Calculating the exact value of the topological entropy of a flow is complicated, but a method to estimate it using braids was described in \citet{Mouss}. A numerical scheme was devised and detailed in \citet{Thiff2010} and a further version of the code developed by Jean-Luc Thiffeault at the University of Wisconsin, Madison. It is a MATLAB package called \normalfont{braidlab}, which is freely available at \normalfont{http://www.math.wisc.edu/\~{}jeanluc/}. This is the package used for this work. Here we give a brief description of the principal of the technique. \\
We now elaborate on the method behind the calculations. As mentioned, when trajectories in the flow are plotted as a function of time, a braid diagram is obtained. Each intersection can be described as a braid generator and labelled according to which trajectories are involved and the sign of the crossing (positive or negative twist). In this way we can build an algebraic representation of a braid, consisting of the braid generators. We can estimate the entropy of the entire flow by calculating the entropy of this braid. The more particle trajectories we track, the closer we get to the true flow entropy. \\
Now, this braiding sequence can be applied to some random material line inserted into the flow. As the particles move around the material line will be stretched and folded accordingly - the higher the rate of stretching, the higher the entropy. We can calculate the number of intersections of the material line with the real axis as it is distorted by the action of the braid. A better estimation can be obtained by taking several different sets of trajectories, computing the intersections, and taking averages in individual time intervals. Plotting these values at the recorded times and taking the gradient of the best fit line gives the entropy estimate. \\
We found that the estimate also improved as the number of periods the driver ran for was increased, therefore giving longer braids to work with and an entropy per period. However the user can run into problems where braids become very long, and repeatedly iterating and trying to calculate intersection numbers can result in attainment of the computational limit. We encountered this problem in our work, and so instead measured the complexity of the braid, another function included in the \normalfont{braidlab} package. This does not repeatedly iterate like the entropy function. Dividing this quantity by the number of periods the braid has been driven for, when this number is large enough, is an alternative method of obtaining the entropy estimate, and is the one we employed. \\

\begin{figure}[ht]
\includegraphics[width=0.5\textwidth]{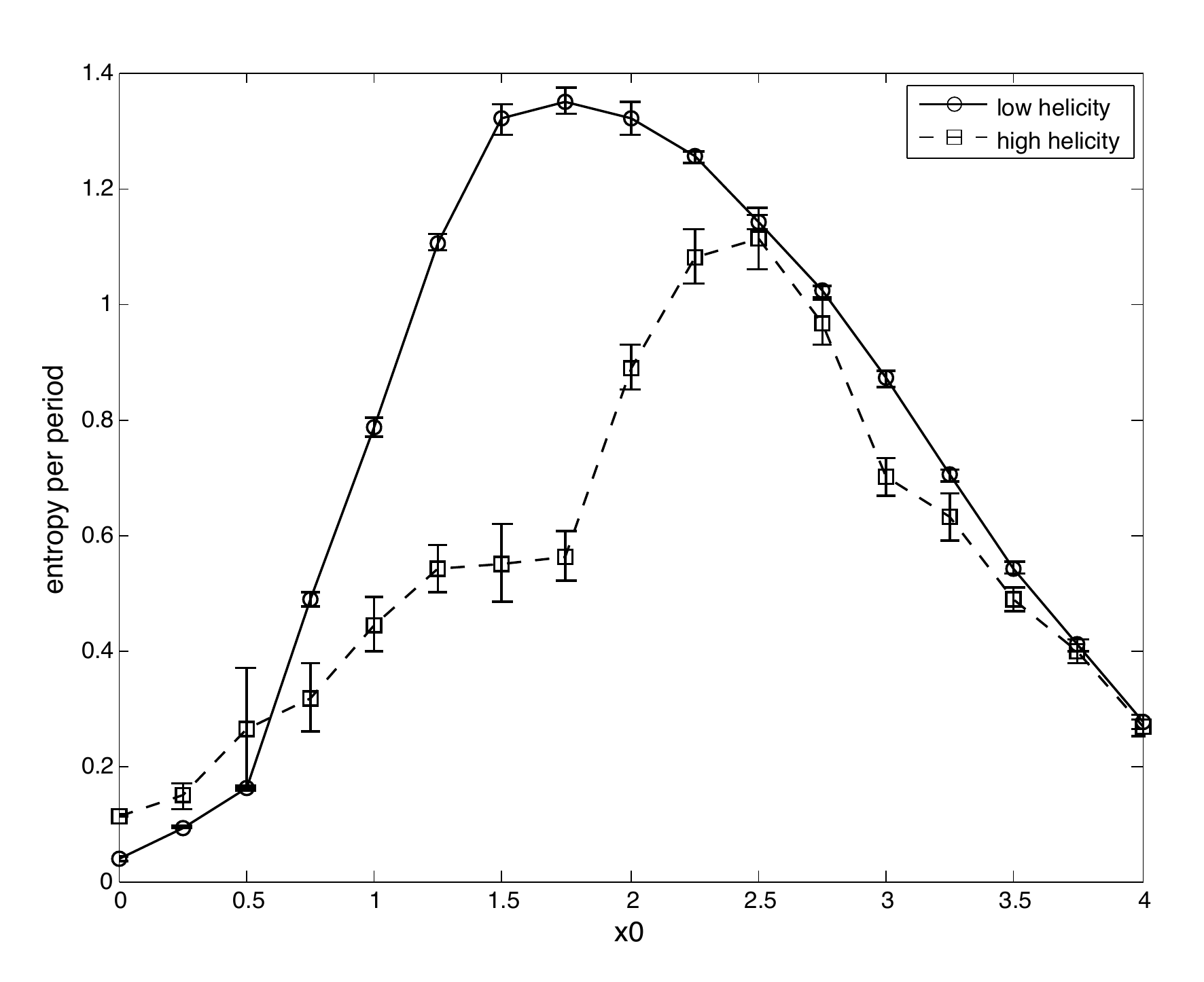}
\caption{Topological entropy calculated for both helicity cases for varying $x_0$. \label{topent}.}
\end{figure}

 Figure \ref{topent} illustrates that the entropy in both helicity cases reaches a peak value as $x_0$ increases from $0$ and then starts to decrease past a critical value. The high helicity, coherent case generally has lower values. Error bars correspond to a $95\%$ confidence interval about a sample mean, sample size 5.\\

\end{document}